\documentclass[a4paper,11pt]{article}
\usepackage{graphicx,epsfig}
\usepackage{fancyhdr,fancybox,float}
\usepackage[title]{appendix}
\usepackage{indentfirst}
\usepackage{verbatim}
\usepackage[sort&compress,numbers]{natbib}
\usepackage{geometry}
\usepackage{extarrows,chemarrow,xypic}
\usepackage{color}
\usepackage[small]{caption2}
\usepackage{microtype}
\DisableLigatures[f]{encoding = *, family = *}

\usepackage{times}                   

\usepackage{amssymb}                 
\usepackage{amsthm}
\usepackage{mathrsfs}                
\usepackage{bbm}
\usepackage{hyperref}                

\newcommand{\paperfont}{\fontsize{11pt}{1.2\baselineskip}\selectfont}
\pagebreak[4]

\begin{document}
	
\theoremstyle{definition}
\makeatletter
\thm@headfont{\bf}
\makeatother
\newtheorem{theorem}{Theorem}[section]
\newtheorem{definition}[theorem]{Definition}
\newtheorem{lemma}[theorem]{Lemma}
\newtheorem{proposition}[theorem]{Proposition}
\newtheorem{corollary}[theorem]{Corollary}
\newtheorem{remark}[theorem]{Remark}
\newtheorem{example}[theorem]{Example}
\newtheorem{assumption}[theorem]{Assumption}

\lhead{}
\rhead{}
\lfoot{}
\rfoot{}

\renewcommand{\refname}{References}
\renewcommand{\figurename}{Figure}
\renewcommand{\tablename}{Table}
\renewcommand{\proofname}{Proof}

\newcommand{\diag}{\mathrm{diag}}
\newcommand{\tr}{\mathrm{tr}}
\newcommand{\Enum}{\mathbb{E}}
\newcommand{\Pnum}{\mathbb{P}}
\newcommand{\Rnum}{\mathbb{R}}
\newcommand{\Cnum}{\mathbb{C}}
\newcommand{\Znum}{\mathbb{Z}}
\newcommand{\Nnum}{\mathbb{N}}
\newcommand{\abs}[1]{\left\vert#1\right\vert}
\newcommand{\set}[1]{\left\{#1\right\}}
\newcommand{\norm}[1]{\left\Vert#1\right\Vert}
\newcommand{\bx}{\mathbf{x}}

\title{\textbf{Some mathematical aspects of Anderson localization: boundary effect, multimodality, and bifurcation}}
\author{Chen Jia$^{1}$,\;\;\;Ziqi Liu$^{2}$,\;\;\;Zhimin Zhang$^{1,3,*}$\\
\footnotesize $^1$ Beijing Computational Science Research Center, Beijing 100193, China \\
\footnotesize $^2$ Department of Mathematical Sciences, Tsinghua University, Beijing 100084, China \\
\footnotesize $^3$ Department of Mathematics, Wayne State University, Detroit, Michigan 48202, U.S.A.\\
\footnotesize Email: zmzhang@csrc.ac.cn}
\date{}
\maketitle
\thispagestyle{empty}

\paperfont

\begin{abstract}
Anderson localization is a famous wave phenomenon that describes the absence of diffusion of waves in a disordered medium. Here we generalize the landscape theory of Anderson localization to general elliptic operators and complex boundary conditions using a probabilistic approach, and further investigate some mathematical aspects of Anderson localization that are rarely discussed before. First, we observe that under the Neumann boundary condition, the low energy quantum states are localized on the boundary of the domain with high probability. We provide a detailed explanation of this phenomenon using the concept of extended subregions and obtain an analytical expression of this probability in the one-dimensional case. Second, we find that the quantum states may be localized in multiple different subregions with high probability in the one-dimensional case and we derive an explicit expression of this probability for various boundary conditions. Finally, we examine a bifurcation phenomenon of the localization subregion as the strength of disorder varies. The critical threshold of bifurcation is analytically computed based on a toy model and the dependence of the critical threshold on model parameters is analyzed. \\

\noindent
\textbf{Keywords}: landscape; spectrum; eigenvalue; eigenmode; eigenfunction; elliptic operator; Schr\"{o}dinger operator; confinement \\

\end{abstract}

\section{Introduction}
Anderson localization is a general wave phenomenon that applies to the transport of electromagnetic waves, quantum waves, spin waves, acoustic waves, etc \cite{anderson1958absence, thouless1974electrons, abrahams1979scaling, lee1985disordered, john1987strong, akkermans1988theoretical, kuhn2005localization, billy2008direct, roati2008anderson}. In particular, it is responsible for the metal-insulator transition in disordered alloys. Due to its importance and universality, it has fascinated scientists and mathematicians in many different areas and produced a huge body of literature in the past 60 years \cite{abrahams2001metallic, evers2008anderson, lagendijk2009fifty}. Since Anderson's seminal work \cite{anderson1958absence}, it is known that a Schr\"{o}dinger operator with a lattice potential can produce a strongly localized quantum state, provided that the degree of randomness (disorder) in the lattice is sufficiently large. This phenomenon has now been experimentally demonstrated
in optic and electromagnetic systems \cite{laurent2007localized, sapienza2010cavity, riboli2011anderson}.

One of the most puzzling aspects of Anderson localization is the strong spatial confinement of quantum states of the system, i.e. the ability to maintain standing waves in a very restricted subregion of the domain, with their amplitudes decaying exponentially at long range even in the absence of confining force or potential. Recently, Filoche and Mayboroda \cite{filoche2012universal} proposed a universal mechanism for Anderson localization. This mechanism reveals that in any vibrating system, there exists a hidden landscape of localization that partitions the original domain into many weakly coupled subregions. The boundaries of these subregions correspond to the valley lines of the landscape. The height of the landscape along its valley lines determines the strength of coupling between the subregions. This theory allows one to predict the localization behavior, including the geometry of the confining subregions, the energy of the quantum states, and the critical energy above which one can expect fully delocalized (conducting) states to appear. Moreover, by constructing an Agmon distance, it has been shown \cite{arnold2016effective, arnold2019localization, arnold2019computing} that the inverse of the localization landscape can be interpreted as an effective confining potential that is responsible for the exponential decay of the localized quantum states even far from its main confining subregion.

Despite more than half a century of research, many issues on the exact mechanism of Anderson localization still remain open \cite{abrahams2001metallic, evers2008anderson, lagendijk2009fifty}. In the present paper, we consider Anderson localization of quantum states of a Schr\"{o}dinger equation with a disordered lattice potential under Neumann and Robin boundary conditions. Using a probabilistic approach that represents the solution of a second-order elliptic equation in terms of a reflecting diffusion, we extend the concepts of localization landscape and its valley lines to general non-symmetric second-order elliptic operators and arbitrary boundary conditions, and we prove that the amplitude of any vibrational eigenmode can be controlled effectively by the landscape. The probabilistic approach is also applied to study the limit localization behavior of quantum states when the strength of disorder tends to infinity.

More importantly, we investigate some novel mathematical aspects of Anderson localization that are seldom discussed in the literature. First, we observe that under the Neumann and Robin boundary conditions, the low energy quantum states are localized on the boundary of the domain with relative high probability. Moreover, we find that compared to the one-dimensional case, the eigenmodes of a two-dimensional system are more likely to be localized on the boundary. These boundary effects are analyzed systematically, which is the first main contribution of the present paper. In addition, we find that the quantum states of a one-dimensional system are often localized in multiple different subregions simultaneously and thus exhibit multiple peaks. We study the multimodal phenomenon in detail, which is the second main contribution of this paper.

Finally, we find that under different strengths of disorder, the quantum states may be localized in completely different subregions. When the disorder is large, the first eigenmode tends to be localized in the largest low potential subregion with strong symmetry. However, when the disorder is small, the first eigenmode tends to be localized in the union of some low potential subregions with some potential barriers between them. As a result, the localization behavior of the system may undergo a bifurcation as the strength of disorder varies. We study such bifurcation phenomenon in detail based on a toy model, which is the third main contribution of this article.

The structure of the present paper is organized as follows. In Section \ref{model}, we extend the concepts of localization landscape and valley lines to general non-symmetric elliptic operators and general boundary conditions, prove the key Filoche-Mayborodaca inequality using a probabilistic approach, and study the limit localization behavior when the strength of disorder is large. In Section \ref{boundary}, we examine the boundary effect of Anderson localization under Neumann and Robin boundary conditions, clarify the reason why the low energy quantum states in high dimensions are more likely to be localized on the boundary of the domain using the concept of extended subregions. In Section \ref{multimodality}, we reveal why the eigenmodes may be localized in multiple different subregions simultaneously and compute the probability of this multimodal phenomenon in the one-dimensional case for various boundary conditions. In Section \ref{transition}, we investigate the bifurcation of the confining subregion as the strength of disorder varies based on a toy model, reveal the essence of the bifurcation phenomenon, and obtain the equation satisfied by the critical threshold of bifurcation. We conclude in Section \ref{conclusion}.

\section{Localization landscape and limit behavior}\label{model}

\subsection{Model}
Here we consider high-dimensional \emph{Anderson localization} for the quantum states of the following stationary Schr\"{o}dinger equation with a Dirichlet, Neumann, or Robin boundary condition (Anderson's original work \cite{anderson1958absence} considers a tight-binding model which can be viewed as the discretization of the continuous model studied here):
\begin{equation}\label{anderson}
\left\{
\begin{split}
& Hu = \lambda u,\quad\textrm{in}\;\Omega, \\
& g \frac{\partial u}{\partial n} + h u = 0,\quad\textrm{on}\;\partial\Omega,
\end{split}
\right.
\end{equation}
where $H = -\triangle + K V$ is the Hamiltonian with $V = V(x)$ being a disordered potential, $\lambda>0$ is an energy level (eigenvalue), $u = u(x)$ is the associated quantum state (eigenmode), and $K > 0$ is the strength of disorder. The domain $\Omega = (0,1)^d$ is the $d$-dimensional unit hypercube with each side being divided uniformly into $N$ intervals. In this way, the hypercube $\Omega$ is divided into $N^d$ smaller hypercubes of the same size. In each small hypercube $\Omega_k$, the potential $V$ is a constant with its value being sampled from a given probability distribution, which is often chosen as the Bernoulli distribution
\begin{equation}\label{bernoulli}
\Pnum(V|_{\Omega_k} = 0) = 1-p, \quad \Pnum(V|_{\Omega_k} = 1) = p,
\end{equation}
or the uniform distribution
\begin{equation*}
\Pnum(a \leq V|_{\Omega_k} \leq b) = \frac{1}{b-a}, \quad 0 \leq a < b \leq 1.
\end{equation*}
The values of $V$ in these small hypercubes are independent of each other. The boundary condition of the model is rather general with $g \geq 0$ being a given constant, $h = h(x) \geq 0$ being a given function on $\partial \Omega$, and $n = n(x)$ being the unit outward pointing normal vector field on $\partial \Omega$. If $g = 0$ and $h = 1$, then it reduces to the Dirichlet boundary condition; if $g = 1$ and $h = 0$, then it reduces to the Neumann boundary condition; if $g = 1$ and $h \neq 0$, then it is the Robin boundary condition.

Note that most previous papers impose the Dirichlet boundary condition in the study of Anderson localization \cite{filoche2012universal}. This is equivalent to trapping a particle by imposing an infinite potential outside the relevant domain. Here we also consider other boundary conditions due to the following reason. Recall that for the time-dependent Schr\"{o}dinger equation $i\partial_t\phi = -H\phi$, the probability density of the particle is defined as $\rho = |\phi|^2 = \phi\bar\phi$ and the probability current $j = j(x)$ of the particle is defined as
\begin{equation*}
j = i(\phi\nabla\bar\phi-\bar\phi\nabla\phi),
\end{equation*}
where $\bar\phi$ is the complex conjugate of $\phi$. The probability density $\rho$ and the probability current $j$ are linked by the continuity equation $\partial_t\rho+\nabla\cdot j = 0$. Note that the Dirichlet, Neumann, and Robin boundary conditions can all guarantee that
\begin{equation*}
j\cdot n = i\left(\phi\frac{\partial\bar\phi}{\partial n}-\bar\phi\frac{\partial\phi}{\partial n}\right)
= 0,\;\;\;\textrm{on}\;\partial\Omega,
\end{equation*}
which means that there is no net probability current across the boundary. Hence imposing the Neumann boundary condition can confine the system within the domain without an infinite potential well, as opposed to the Dirichlet boundary condition. This is the case of perfect reflection on the boundary (see Section 5.2 of \cite{thaller2007visual} for a detailed explanation). The Neumann boundary condition turns out to be very useful in the $R$-matrix theory of scattering \cite{pichugin2001effective, lee2010r} and the theory of quantum graphs \cite{harrell2020localization}. The Robin boundary condition is also considered here since it builds a bridge between the Dirichlet and Neumann boundary conditions.

In fact, the above model can be generalized to a more complicated model as follows:
\begin{equation}\label{eigenproblem}
\left\{
\begin{split}
& -Lu+KVu = \lambda u, \quad\textrm{in}\;\Omega, \\
& \quad g \frac{\partial u}{\partial n} + h u = 0, \quad \textrm{on} \; \partial \Omega,
\end{split}
\right.
\end{equation}
where
\begin{equation}\label{operator}
L = \frac{1}{2}\sum_{i,j=1}^{d}a^{ij}(x)\partial_{ij}+\sum_{i=1}^{d}b^i(x)\partial_i
\end{equation}
is an arbitrary second-order elliptic operator, $V = V(x)$ is an arbitrary random potential, and $\Omega \subset \Rnum^d$ is an arbitrary bounded domain. Note that the operator $L$ may be non-symmetric. Mathematically, the operator $L$ is called symmetric if there exists a smooth function $\rho = \rho(x)$ such that
\begin{equation*}
(Lf,g)_{\rho} = (f,Lg)_{\rho},\;\;\;\forall f,g\in C_c^\infty(\Omega),
\end{equation*}
where $C_c^{\infty}(\Omega)$ is the space of all smooth functions on $\Omega$ with compact supports and
\begin{equation*}
(f,g)_{\rho} = \int_{\Omega}f(x)\bar{g}(x)\rho(x)dx.
\end{equation*}
We emphasize that here the symmetry of the operator $L$ should be distinguished from the symmetry of the matrix $A = (a^{ij})_{d \times d}$ of diffusion coefficients. The latter is always symmetric, but the former may not. It is well-known that the operator $L$ is symmetric if and only if the vector field $A^{-1}(2b-\nabla\cdot A)$ is conservative (has a potential function), i.e.
there exists a smooth function $U = U(x)$ such that
\begin{equation*}
A^{-1}(2b-\nabla\cdot A) = -\nabla U,
\end{equation*}
where $A = (a^{ij})$ is the diffusion matrix, $b = (b^i)$ is the drift, and $\nabla\cdot A$ is a vector field whose $i$th component is given by $(\nabla\cdot A)^i = \partial_j a^{ij}$ \cite{jiang2004mathematical, ge2021martingale}. Clearly, the Laplace operator $\triangle$ is symmetric. The eigenvalues of a symmetric operator must be all real numbers, while the eigenvalues of a non-symmetric operator may be complex numbers. In the following, we shall define the localization landscape and its valley lines for the model \eqref{eigenproblem}.

\subsection{Localization landscape}
A recent theory \cite{filoche2012universal} has shown that under the Dirichlet boundary condition ($g = 0$ and $h = 1$), the spatial location of the localized quantum states of the eigenvalue problem \eqref{anderson} can be predicted by the solution of an associated Dirichlet problem
\begin{equation}\label{landDirichlet}
\left\{
\begin{split}
& -\triangle w + K V w = 1, \quad \textrm{in} \; \Omega, \\
& \quad w = 0, \quad \textrm{on} \; \partial \Omega,
\end{split}
\right.
\end{equation}
where the solution $w = w(x)$ is called the \emph{localization landscape}. In fact, the theory in \cite{filoche2012universal} was developed for symmetric elliptic operators and the Dirichlet boundary condition using the technique of the Green function. The landscape theory was further developed in \cite{steinerberger2017localization} using a probabilistic approach and generalized in \cite{arnold2019computing} to the Neumann boundary condition. Here we extend the concept of localization landscape to general non-symmetric elliptic operators and more complicated boundary conditions using a different probabilistic approach.

\begin{figure}
\centering\includegraphics[width=\linewidth]{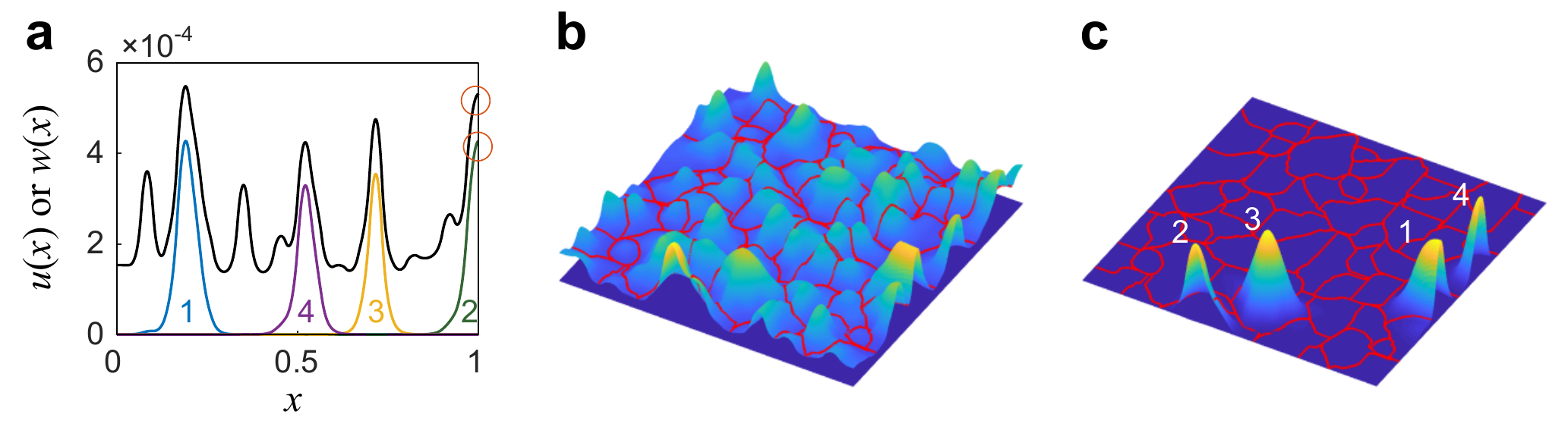}
\caption{\textbf{Localization landscape and valley lines under Neumann boundary conditions.} (a) Localization landscape $w$ (black curve) and the first four eigenmodes $u/\lambda$ (colored curves) for a one-dimensional problem under the Neumann boundary condition. The colored numbers show the order of the eigenmodes. Note that both the second eigenmode $u$ and the landscape $w$ attain their local maxima at $x = 1$ (red circles) and also have zero derivatives at $x = 1$ (the zero derivatives for the two functions at the right boundary are not apparent because the localized peaks are very sharp). The parameters are chosen as $K = 8000$ and $N = 30$. (b) Landscape and valley lines (red curves) for a two-dimensional problem under the Neumann boundary condition. The parameters are chosen as $K = 8000$ and $N = 20$. (c) Valley lines and the first four eigenmodes for the system in (b). In (a)-(c), the operator $L$ is chosen to be the laplace operator and the potential $V$ in each small hypercube is sampled from the uniform distribution over $[0,1]$.}\label{fig1}
\end{figure}

The key to our approach is to find the probabilistic representation for the solution of the eigenvalue problem \eqref{eigenproblem}. For the Dirichlet boundary condition, the solution can be represented by the expectation of a functional of a Brownian motion \cite{steinerberger2017localization}. However, for Neumann and Robin boundary conditions, the solution can no longer be represented by a Brownian motion, but should be represented by a reflecting diffusion. To see this, recall that for a bounded domain $\Omega \subset \Rnum^d$ with a unit outward pointing normal vector field $n = n(x)$ on $\partial \Omega$, the operator $L$ given in \eqref{operator} is the infinitesimal generator of a reflecting diffusion $X = (X_t)_{t \geq 0}$ with drift $b = (b^i)$ and diffusion matrix $a = (a^{ij})$. This reflecting diffusion is the solution to the Skorokhod stochastic differential equation
\begin{equation}\label{reflecting}
dX_t = b(X_t) dt + a^{1/2}(X_t) d B_t - g n(X_t) d F_t, \quad X_0 = x\in\Omega,
\end{equation}
where $B = (B_t)_{t\geq 0}$ is a $d$-dimensional standard Brownian motion and $F_t$ is a continuous nondecreasing process that increases only when $X_t \in \partial \Omega$ \cite{1998Diffusions}. In particular, if $L = \triangle$ is the Laplace operator, then the solution of \eqref{reflecting} is a reflecting Brownian motion. It can be proved that when $g > 0$, the reflecting diffusion $X_t$ can never exit the domain $\Omega$; once $X_t$ touches $\partial \Omega$, it will be reflected into $\Omega$ again due to the effect of the random force $F_t$ \cite{1998Diffusions}. With the aid of the reflecting diffusion, it can be shown that the solution of the problem \eqref{eigenproblem} has the following probabilistic representation (see Appendix \ref{appA} for the proof):
\begin{equation}\label{probeigen}
u(x) = \lambda \Enum_x \int_{0}^{\infty} u(X_t) e^{- \int_{0}^{t} h(X_s) d F_s - \int_{0}^{t} K V(X_s) ds} dt, \quad \; x \in \Omega,
\end{equation}
where $\Enum_x$ denotes the conditional expectation given that $X_0 = x$. For the problem \eqref{eigenproblem}, we define its localization landscape $w = w(x)$ to be the solution of the following boundary value problem:
\begin{equation}\label{landscape}
\left\{
\begin{split}
& - L w + K V w = 1 \quad \textrm{in} \; \Omega, \\
& \quad g \frac{\partial w}{\partial n} + h w = 0 \quad \textrm{on} \; \partial \Omega.
\end{split}
\right.
\end{equation}
Note that the boundary value problem \eqref{landscape} is obtained by setting the right-hand side of the eigenvalue problem \eqref{eigenproblem} to be $1$. Similarly, the landscape $w$ also has a probabilistic representation using reflecting diffusion, which is given by (see Appendix \ref{appA} for the proof)
\begin{equation}\label{probland}
w(x) = \Enum_x \int_{0}^{\infty} e^{- \int_{0}^{t} h(X_s) d F_s - \int_{0}^{t} KV(X_s) ds} dt, \quad \; x \in \Omega.
\end{equation}
The probabilistic representations \eqref{probeigen} and \eqref{probland} are closely related. If we normalize the eigenmode $u$ such that $\|u\|_\infty = 1$, then we obtain the \emph{Filoche-Mayboroda inequality}
\begin{equation}\label{FMinequality}
|u(x)| \leq |\lambda| w(x), \quad \; x \in \Omega.
\end{equation}
This inequality shows that when the energy level $\lambda$ is not large, the eigenmode $u$ must be small at those points where the landscape $w$ is small. Hence the low energy eigenmodes must be localized to those subregions where the landscape is large. To see this, we illustrate the graphs of the landscape $w$ and the first four eigenmodes $u/\lambda$ for a one-dimensional problem when $K$ is large (Fig. \ref{fig1}(a)). It can be seen that the inequality \eqref{FMinequality} indeed provides an accurate upper bound for the eigenmodes. Furthermore, the spatial locations of the confining subregions of these eigenmodes are perfectly predicted by the peak positions of the landscape. Recent numerical and theoretical studies \cite{arnold2019computing, chenn2021approximating} have shown that the order of the eigenmodes is closely related to the height of the peaks of the landscape. This phenomenon was observed in our simulations in Fig. \ref{fig1}(a), where the first four eigenmodes are localized around the heighest four peaks of the landscape.

\subsection{Valley lines}
In the one-dimensional case, the local minima of the landscape $w$ divide the domain $\Omega$ into many subintervals (Fig. \ref{fig1}(a)). Such subintervals give possible spatial locations where the eigenmodes can be localized. In the two-dimensional case, the confining subregions are determined by the \emph{valley lines} of the landscape \cite{filoche2012universal}. In fact, the valley lines are defined as the lines of steepest descent, starting from the saddle points of the landscape and going to its local minima. In this way, the valley lines separate the domain $\Omega$ into many subregions. Along the valley lines, the values of $w$ are expected to be small. From the inequality \eqref{FMinequality}, the values of $u$ are also small along the valley lines and thus the eigenmodes must be localized within the subregions enclosed by these lines.
Following \cite{filoche2012universal, arnold2016effective}, we use the watershed algorithm proposed in \cite{Soille1990Determining} to compute the valley lines numerically. In higher dimensions, valley lines will become hypersurfaces and the results are similar.

Fig. \ref{fig1}(b),(c) illustrate the landscape, eigenmodes, and valley lines for a two-dimensional problem under the Neumann boundary condition. It is clear that the valley lines of the landscape separate the domain into many subregions and each eigenmode is exactly located in one of these subregions. Under the Dirichlet boundary condition, the eigenmodes must be located in the interior of the domain since they vanish on the boundary \cite{filoche2012universal}. However, under the Neumann boundary condition, some eigenmodes may be localized on the boundary (see the first, second, and fourth ones in Fig. \ref{fig1}(c)). The boundary values of these eigenmodes are large and their local maxima may even appear on the boundary. This phenomenon will never appear for the Dirichlet boundary condition.

Here the eigenvalue problem \eqref{eigenproblem} and the boundary value problem \eqref{landscape} are solved numerically by using the spectral element method instead of the classical finite difference method (FDM) or finite element method (FEM). In the spectral element method, the solution of the associated partial differential equation (PDE) is approximated by piecewise high-order polynomials and the Legendre polynomials are used as a basis to approximate the solution in each small hypercube $\Omega_k$. The degree of the Legendre polynomials is chosen to be $10$ in the one-dimensional case and $6$ in the two-dimensional case. In fact, classical FDM or FEM only have second-order convergence with respect to the mesh size, while the spectral element method can achieve exponential convergence with respect to the degree of the polynomials. Therefore, the spectral element method can obtain higher numerical accuracy with less computational costs than the other two methods.

\subsection{Limit behavior for large $K$}
Next we focus on the limit behavior of the eigenmodes when the strength $K$ of disorder is very large. For simplicity, we assume that the potential $V$ restricted to each small hypercube $\Omega_k$ is sampled from the \emph{Bernoulli distribution} given in \eqref{bernoulli}, which can only take the value of $0$ or $1$. To proceed, let
\begin{equation*}
D = \{x \in \Omega: V(x) = 0\}
\end{equation*}
denote the set of points at which the potential vanishes. We first consider the behavior of the eigenmodes outside $D$. Let $x\in\Omega$ be the initial position of the reflecting diffusion $X_t$ which solves the Skorokhod equation \eqref{reflecting}. When $x \notin D$, we have $V(X_s) = 1$ when $s$ is small, which implies that
\begin{equation*}
\int_{0}^{t} V(X_s) ds > 0, \quad \; t > 0.
\end{equation*}
It thus follows from \eqref{probeigen} and \eqref{probland} and the dominated convergence theorem that
\begin{equation}\label{largeK}
\lim_{K \rightarrow \infty} w(x) = \lim_{K \rightarrow \infty} u(x) = 0, \quad \; x \notin D,\;x \in \Omega.
\end{equation}
This shows that the landscape and eigenmodes must vanish outside $D$ in the limit of $K\rightarrow\infty$. In other words, the eigenmodes can only be localized in the region where the potential attains its minimum. This coincides with the intuition that the quantum states tend to be localized in the region with low potential energy.

We next focus on the behavior of the eigenmodes inside $D$. Clearly, $D$ can be decomposed as the disjoint union of several connected components (subregions), i.e. $D = D_1 \cup D_2 \cup \cdots \cup D_M$. Since the potential $V$ vanishes in $D$, in the limit of $K\rightarrow\infty$, the eigenmode $u$ must be the solution to the following local eigenvalue problem in each subregion $D_k$:
\begin{equation}\label{subregion}
\left\{
\begin{split}
& - L u = \lambda u \quad \textrm{in}\;\;D_k, \\
& \quad g \frac{\partial u}{\partial n} + h u = 0 \quad \textrm{on}\;\;\partial D_k \cap \partial \Omega, \\
& \quad u = 0 \quad \textrm{on}\;\;\partial D_k \setminus \partial \Omega.
\end{split}
\right.
\end{equation}
Therefore, when $K$ is very large, the spectrum of the Hamiltonian $H = -L+KV$ is composed of the local eigenvalues of the operator $-L$ in each subregion $D_k$. If an eigenvalue of $H$ coincides with one of the local eigenvalues of $-L$ in $D_k$, then the corresponding eigenmode will be localized in $D_k$; conversely, if an eigenvalue of $H$ coincides with neither one of the local eigenvalues of $-L$ in $D_k$, then the corresponding eigenmode will not be localized in $D_k$. Hence the limits in \eqref{largeK} enable us to decompose the eigenvalues problem \eqref{eigenproblem} with random potential in the whole domain $\Omega$ into some local eigenvalue problems \eqref{subregion} with zero potential in the subregions $D_k$. When $K$ is not very large, the landscape and eigenmodes are small but not zero on $\partial D_k \setminus \partial \Omega$ and thus the domain $\Omega$ can be divided into the many weakly coupled subregions \cite{filoche2012universal}.

\begin{figure}
\centering
\includegraphics[width=\linewidth]{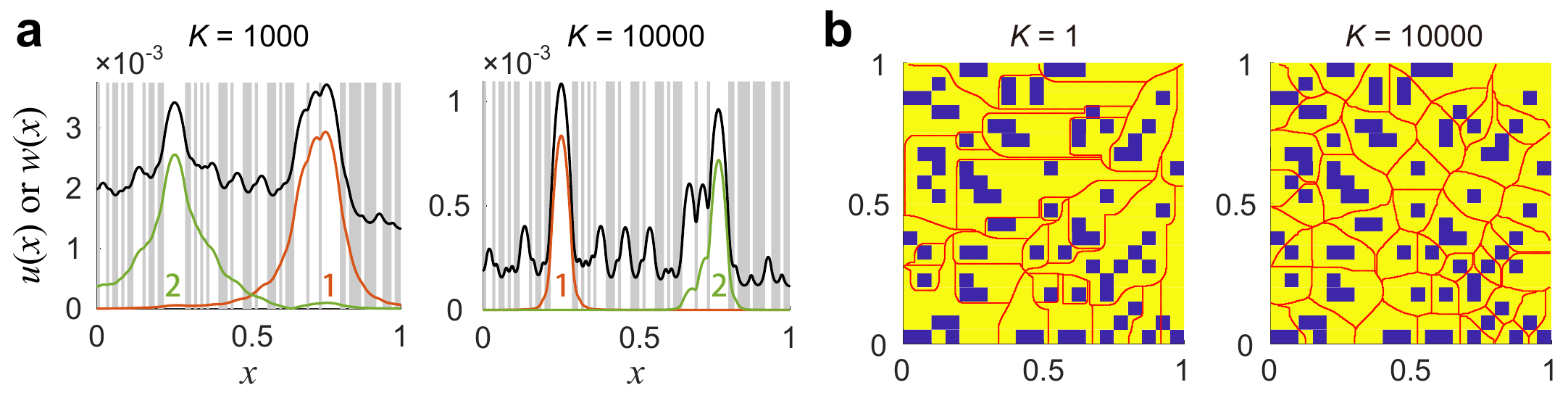}
\caption{\textbf{Influence of the strength $K$ of disorder on Anderson localization}. (a) Potential (grey and white regions), landscape (black curve) and the first two eigenmodes (colored curves) for a one-dimensional problem under the Neumann boundary condition as $K$ varies. The grey parts show the subregions with $V=1$ and the white parts show the subregions with $V=0$. The parameters are chosen as $N=100$ and $p=0.5$. (b) Potential (yellow and purple regions) and valley lines (red curves) for a two-dimensional problem under the Neumann boundary condition as $K$ varies. The yellow parts show the subregions with $V = 1$ and the purple parts show the subregions with $V = 0$. The parameters are chosen as $N=20$ and $p=0.8$.}
\label{fig2}
\end{figure}

To gain a deeper insight, we depict the potential, landscape, and eigenmodes for a one-dimensional problem under different values of $K$ (Fig. \ref{fig2}(a)). When $K$ is large, the landscape and eigenmodes are only localized in the region where the potential vanishes. However, this is not the case when $K$ is relatively small. Fig. \ref{fig2}(b) illustrate the potential and valley lines for a two-dimensional problem under different values of $K$. When $K$ is small, the confining subregions enclosed by the valley lines may include many connected components of $D$. However, for sufficiently large $K$, the connected components of $D$ are exactly separated by valley lines.

\section{Boundary effect}\label{boundary}
In the previous discussion, we develop our theory for the general model \eqref{eigenproblem}. In the following, we only focus on the simpler model \eqref{anderson}. For simplicity, we assume that the potential $V$ restricted to each small hypercube $\Omega_k$ is Bernoulli distributed.

\subsection{Localization on the boundary}
We have seen that under the Neumann or Robin boundary condition, the low energy quantum states are very likely be localized on the boundary (see the second eigenmode in Fig. \ref{fig1}(a) and the first, second, and fourth eigenmodes in Fig. \ref{fig1}(c)). Such phenomenon will never take place for the Dirichlet boundary condition. When an eigenmode is localized on the boundary, its boundary value must be very large and the local maxima of the eigenmode may even appear on the boundary. In what follows, each eigenmode is always normalized such that $\|u(x)\|_\infty = 1$. To better characterize the boundary effect, we introduce the following definition. In the one-dimensional case, we say that an eigenmode $u$ is localized on the boundary if it satisfies
\begin{equation}\label{probbound}
\max\{|u(0)|, |u(1)|\} > 0.5.
\end{equation}
In the two-dimensional case, similarly, we say that an eigenmode $u$ is localized on the boundary if it satisfies
\begin{equation}\label{probedge}
\max_{x \in \partial \Omega} |u(x)| > 0.5,
\end{equation}
and we say that it is localized in the corner if it satisfies
\begin{equation}\label{probcorner}
\max\{|u(0,0)|, |u(0,1)|, |u(1,1)|, |u(1,0)|\} > 0.5.
\end{equation}
Since the potential $V$ is stochastic, the eigenmodes may or may not be localized on the boundary. For simplicity, we only consider the first eigenmode. The probability for the first eigenmode to be localized on the boundary is denoted by $P_b$ and the probability for the first eigenmode to be localized in the corner is denoted by $P_c$. These two probabilities are collectively referred to as \emph{boundary probabilities}. Clearly, the boundary probabilities are both $0$ for the Dirichlet boundary condition. However, for the Neumann or Robin boundary condition, these probabilities are strictly positive with their values depending on the parameters $h$, $K$, and $p$, where the function $h$ is chosen to be a constant for simplicity.
\begin{figure}[!htb]
\centering\includegraphics[width=\linewidth]{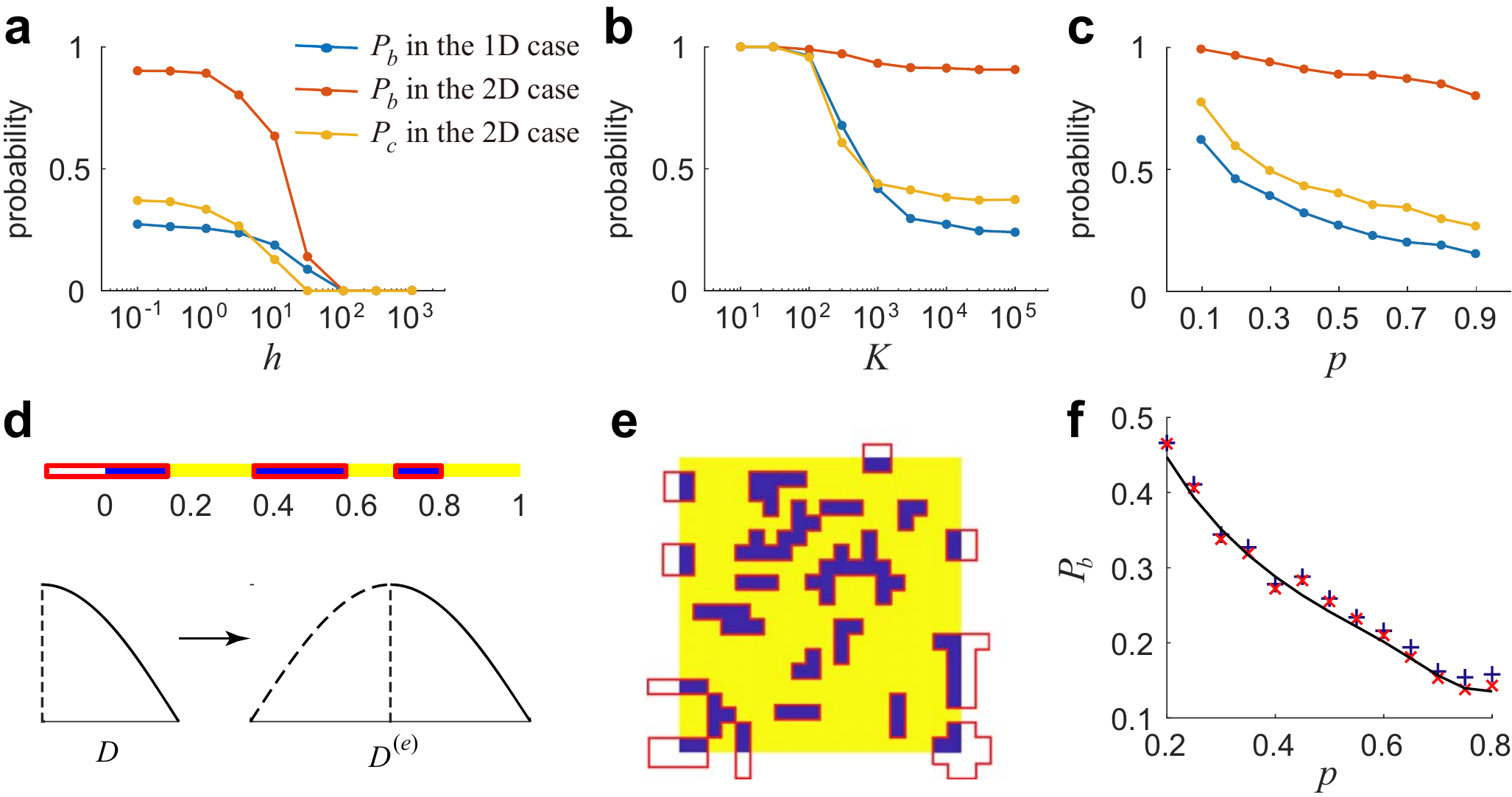}
\caption{\textbf{Boundary effect of Anderson localization under the Robin boundary condition}. (a)-(c) Influence of the parameters $h$, $K$, and $p$ on the boundary probabilities $P_b$ and $P_c$ in the one- and two-dimensional cases. Except the one that is tuned, the other parameters are chosen as $K = 10^3, h = 0.01, p = 0.5$. The parameter $N$ is chosen as $N = 50$ in the one-dimensional case and $N = 15$ in the two-dimensional case. (d) Extended subregions (red boxes) in the one-dimensional case. The yellow parts show the subregions with $V = 1$ and the purple parts show the subregions with $V = 0$. The first eigenmode of the local problem \eqref{regionD} in the subregion $D$ can be viewed as that of the equivalent problem \eqref{regionDe} in the extended subregion $D^{(e)}$. (e) Extended subregions (red boxes) in the two-dimensional case. (f) Dependence of the boundary probability $P_b$ on the parameter $p$ in the one-dimensional case when $h$ is small and $K$ is large. The black curve shows the theoretical prediction given in \eqref{bdprob}, the blue plus signs show the frequency at which the first eigenmode is localized on the boundary, and the red crosses show the frequency at which the longest extended subregion is located on the boundary. Here the frequencies are obtained by generating the random potential 1000 times. The parameters are chosen as $K = 5 \times 10^4, h = 0.01, N = 50$.}
\label{fig3}
\end{figure}

Fig. \ref{fig3}(a)-(c) illustrate the boundary probabilities as $h$, $K$, and $p$ vary under the Robin boundary condition. Here the boundary probabilities $P_b$ and $P_c$ are computed numerically using the Monte Carlo method. Specifically, we generate the random potential $1000$ times, solve the corresponding eigenvalue problems, and then count the frequency that the first eigenmode appears on the boundary or in the corner. The spectral element method enables us to solve the eigenvalue problem thousands of times in an acceptable time. It can be seen that the boundary probabilities decrease with the three parameters in both the one- and two-dimensional cases. When $h$ is small, the first eigenmode is localized on the boundary with relatively high probability. In particular, the probability $P_b$ is exceptionally large in the two-dimensional case. When $h\gg 1$, the Robin boundary condition reduces to the Dirichlet one and thus all boundary probabilities tend to zero (Fig. \ref{fig3}(a)). This explains why the boundary probabilities decrease as $h$ increases.

We next focus on the dependence of the boundary probabilities on $K$. When $K = h = 0$, the problem \eqref{anderson} reduces to the Laplacian eigenvalue problem with the Neumann boundary condition and thus the first eigenmode is a constant. According to the definition, a constant eigenmode is localized on the boundary (since its maximal boundary value is larger than $0.5$). As $K$ increases, the eigenmode may be localized in the interior of the domain and thus the boundary probabilities decrease. When $K\gg 1$, compared to the one-dimensional case, we find that the eigenmodes for a two-dimensional problem are more likely to be localized on the boundary (Fig. \ref{fig3}(b)). The effect of $p$ on the boundary probabilities is much more complicated and will be discussed later.

\subsection{An explanation of the boundary effect}
Based on the simulations in Fig. \ref{fig3}(a)-(c), we have made two crucial observations: (i) when $h$ is small, the eigenmodes are localized on the boundary with relatively high probability; (ii) when $K$ is large, compared to the one-dimensional case, the eigenmodes for a two-dimensional problem are more likely to be localized on the boundary. Here we provide a detailed explanation for these two observations when $h = 0$ (Neumann boundary condition) and $K\gg 1$ (strong disorder).

In fact, the boundary effect can be explained intuitively using the energy functional theory. To see this, recall that for the Dirichlet and Neumann boundary conditions, the eigenmodes of \eqref{anderson} are actually the critical points of the energy functional \cite{thaller2007visual}
\begin{equation}\label{energyfunc}
J(u) = \int_{\Omega}|\nabla u|^2dx+\int_{\Omega}KV|u|^2dx,
\end{equation}
where the first term is the kinetic energy and the second term is the potential energy. When $K\gg 1$ is large, the potential energy is dominant whenever $V\neq 0$ and thus the low energy quantum states must be localized in the region $D = \bigcup_{k=1}^MD_k$ where the potential $V$ vanishes. Hence we only need to focus on the kinetic energy on each subregion $D_k$, i.e.
\begin{equation}
J_k(u) = \int_{D_k}|\nabla u|^2dx.
\end{equation}
Hence the energy functional perspective also guides us to focus on the local eigenvalue problem in each subregion $D_k$. Intuitively, for the Dirichlet boundary condition, the function $u$ must tend to zero on the boundary, which leads to a large $|\nabla u|$ and thus a large kinetic energy $J_k(u)$ when $D_k$ appears on the boundary. For the Neumann boundary condition, however, the function $u$ does not have to be zero on the boundary, which makes the kinetic energy $J_k(u)$ much smaller. This explains why the boundary effect is significant for the Neumann boundary condition.

We next examine the boundary effect in more detail. For simplicity, we first focus on the one-dimensional case. For large $K$, we have shown that the eigenmode of the problem \eqref{anderson} restricted to each subregion $D_k$ is approximately the solution to the local problem \eqref{subregion}, which is a Laplacian eigenvalue problem. The first eigenmode must be localized in the subregion $D_k$ where the local eigenvalue is the smallest. Suppose that $D_k = [x_k, x_k+L]$, where $L$ is the length of $D_k$. When $D_k \cap \partial \Omega = \varnothing$, i.e. $x_k>0$ and $x_k+L<1$, the local problem \eqref{subregion} can be rewritten as
\begin{equation}\label{regionD}
\left\{
\begin{split}
& - u''(x) = \lambda u(x), \quad x \in (x_k, x_k+L), \\
& \quad u(x_k) = u(x_k+L) = 0,
\end{split}
\right.
\end{equation}
which has the Dirichlet boundary condition. The first eigenvalue and eigenmode of the system are given by $\lambda_1 = (\pi/L)^2$ and $u_1(x) = \sin(\pi(x-x_k)/L)$, respectively. Clearly, the longer the subregion $D_k$, the smaller the local eigenvalue. If the original system has the Dirichlet boundary condition, then the first eigenmode must be localized in the longest subregion when $K\gg 1$.

When $D_k \cap \partial \Omega \neq \phi$, i.e. $x_k=0$ or $x_k+L=1$, the local problem \eqref{subregion} has a mixed boundary condition. For simplicity, we only consider the case of $x_k = 0$. In this case, the local problem can be rewritten as
\begin{equation}\label{regionDe}
\left\{
\begin{split}
& - u''(x) = \lambda u(x), \quad x \in (0, L), \\
& \quad u'(0) = u(L) = 0,
\end{split}
\right.
\end{equation}
where the left-hand side of the interval has the Neumann boundary condition and the right-hand side has the Dirichlet one. The first eigenvalue and eigenmode of the system are given by $\lambda_1 = (\pi/2L)^2$ and $u_1(x) = \cos(\pi x/2L)$, respectively. We then make a crucial observation that the first eigenpair of this system can be regarded as that of another system in the extended subregion $D_k^{(e)} = [-L, L]$:
\begin{equation}\label{equivalent}
\left\{
\begin{split}
& - u''(x) = \lambda u(x), \quad x \in (-L, L), \\
& \quad u(-L) = u(L) = 0,
\end{split}
\right.
\end{equation}
which has the Dirichlet boundary condition (see Fig. \ref{fig3}(d) for an illustration of the extended subregion and the first eigenmodes in the original and extended subregions). In other words, when $D_k$ is around the boundary, the first eigenmode of the local system \eqref{regionDe} with mixed boundary condition can be viewed as that of an equivalent system \eqref{equivalent} with Dirichlet boundary condition in the extended subregion $D_k^{(e)}$ which is twice as long as the original subregion $D_k$. Clearly, for a subregion of a fixed length, the local problem \eqref{regionDe} has a smaller eigenvalue if the subregion appears on the boundary since the length of the subregion should be ``doubled". From the energy functional perspective, compared to the Dirichlet boundary condition, the kinetic energy $J_k(u)$ for the Neumann boundary condition is smaller by a factor of $2$ when $D_k$ appears on the boundary (note that the factor may not be $2$ in higher dimensions). This explain why compared to the interior of the domain, the eigenmodes are more likely to be localized on the boundary when $h$ is small and $K$ is large.

The above considerations can also be generalized to higher dimensions. For each subregion $D_k$, we can extend it to another subregion $D_k^{(e)}$ via mirror symmetry along the boundary (Fig. \ref{fig3}(e)). If $D_k$ does not appear on the boundary, then the extended subregion $D_k^{(e)}$ is the same as $D_k$; if $D_k$ appears on the boundary, then $D_k^{(e)}$ is strictly larger than $D_k$. Suppose that $D_k$ appears on the boundary. Then the local problem \eqref{subregion} has the Neumann boundary condition on $\partial D_k \cap \partial \Omega$ and the Dirichlet boundary condition on $\partial D_k \setminus \partial \Omega$. In Appendix \ref{appD}, we have proved that the first eigenmode of the local system with mixed boundary condition is exactly the same as that of an equivalent system with Dirichlet boundary condition in the extended subregion $D_k^{(e)}$. In particular, in the two-dimensional case, if $D_k$ does not lie in the corner, then the extended subregion is twice as large as the original subregion. However, if $D_k$ lies in the corner, since there are two edges enclosing $D_k$, the extended subregion should be reflected along both edges and thus is four times as large as the original subregion. Recall that a Laplacian eigenvalue problem tends to have smaller eigenvalues on a larger domain with strong symmetry. Clearly, if $D_k$ appears on the boundary, especially in the corner, then it has a larger extended subregion which also has strong symmetry (Fig. \ref{fig3}(e)). Therefore, subregions around the boundary tend to have smaller local eigenvalues and thus the eigenmodes are more likely to be localized on the boundary.

With the aid of the concept of extended subregions, we are able to transform a local eigenvalue problem with a mixed boundary condition into an equivalent eigenvalue problem with the Dirichlet boundary condition in an extended subregion. When $h$ is small and $K$ is large, the first eigenmode will be localized in the extended subregion with the smallest Dirichlet eigenvalue. In particular, in the one-dimensional case, the first eigenmode will be localized in the longest extended subregion. If the longest two extended subregions are the same in length, then the first eigenmode may be localized in these two subregions simultaneously. This phenomenon will be discussed in detail later.

Note that in the one-dimensional case, there are at most two subregions around the boundary and the extended subregion is twice as long as the original one. However, in the two-dimensional case, the proportions of subregions that need to be extended is much higher (Fig. \ref{fig3}(e)). Moreover, if a subregion appears in the corner, then its area should be ``magnified" four times. These two facts explain why compared to the one-dimensional case, the eigenmodes for a two-dimensional problem have a much higher probability to be localized on the boundary.

\subsection{Dependence of the boundary probabilities on $p$}
In the previous discussion, we have explained why the boundary probabilities decrease as $h$ and $K$ increase. Here we focus on the dependence of the boundary probabilities on $p$. For simplicity, we only focus on the one-dimensional case for small $h$ and large $K$.

In the one-dimensional case, the domain $\Omega = (0, 1)$ is divided uniformly into $N$ subintervals of the same length $1/N$. In each subinterval, the potential $V$ can only take the value of $0$ or $1$ with $p$ being the probability of $V = 0$ and $q = 1-p$ being the probability of $V = 1$. Then the domain $\Omega$ can be decomposed into many subregions with $V = 0$ and $V = 1$ alternatively (see the white and gray regions in Fig. \ref{fig2}(a)). Intuitively, when $N$ is large, the lengths of these subregions are approximately independent and the number of subintervals included in each subregion with $V = 0$ ($V = 1$) approximately follows a geometric distribution with parameter $p$ ($q$). For small $h$ and large $K$, we have shown that the first eigenmode must be localized in the longest extended subregion with $V = 0$. Therefore, the probability that the first eigenmode is localized on the boundary is approximately equal to the probability that the longest extended subregion with $V = 0$ appears on the boundary, which is given by (see Appendix \ref{appB} for the proof)
\begin{equation}\label{bdprob}
P_b \approx q^2 p^2 \sum_{k=1}^{\infty} q^{k-2} \sum_{n=1}^{k-1} (1 - q^{2 \max\{k-n,n\}-1})^{M-2} + 2 p^2 q \sum_{n=1}^{\infty} (1-q^{2n-1})^{M-1} q^{n-1}.
\end{equation}
This formula reveals a complex relationship between the boundary probability $P_b$ on the parameter $p$ in the one-dimensional case.

To test our theory, we compare the theoretical prediction given in \eqref{bdprob} with numerical simulations where the boundary probability $P_b$ is computed by generating the random potential 1000 times and then solving the associated eigenvalue problem \eqref{anderson} (Fig. \ref{fig3}(f)). Interestingly, while the analytical expression is very complicated, it is in perfect agreement with simulation results. Moreover, we find that $P_b$ decreases with $p$. When $p\approx 0.5$, we have $P_b\approx 0.26$, which means that the first eigenmode has a one in four chance of being localized on the boundary. In the two-dimensional case, the dependence of the boundary probabilities $P_b$ or $P_c$ on the parameter $p$ is similar (Fig. \ref{fig3}(c)), but it is very difficult to obtain their analytical expressions.

\subsection{Boundary effects for other types of random potentials}
Thus far, the boundary effects was examined when the lattice potential $V$ within each hypercube is randomly sampled from a Bernoulli distribution. From Fig. \ref{fig1}, it is clear that the boundary effect is also significant when the random potential $V$ is sampled from a uniform distribution. A natural question is whether the boundary effect is also significant for other types of random potentials.
\begin{figure}[!htb]
\centering\includegraphics[width=\linewidth]{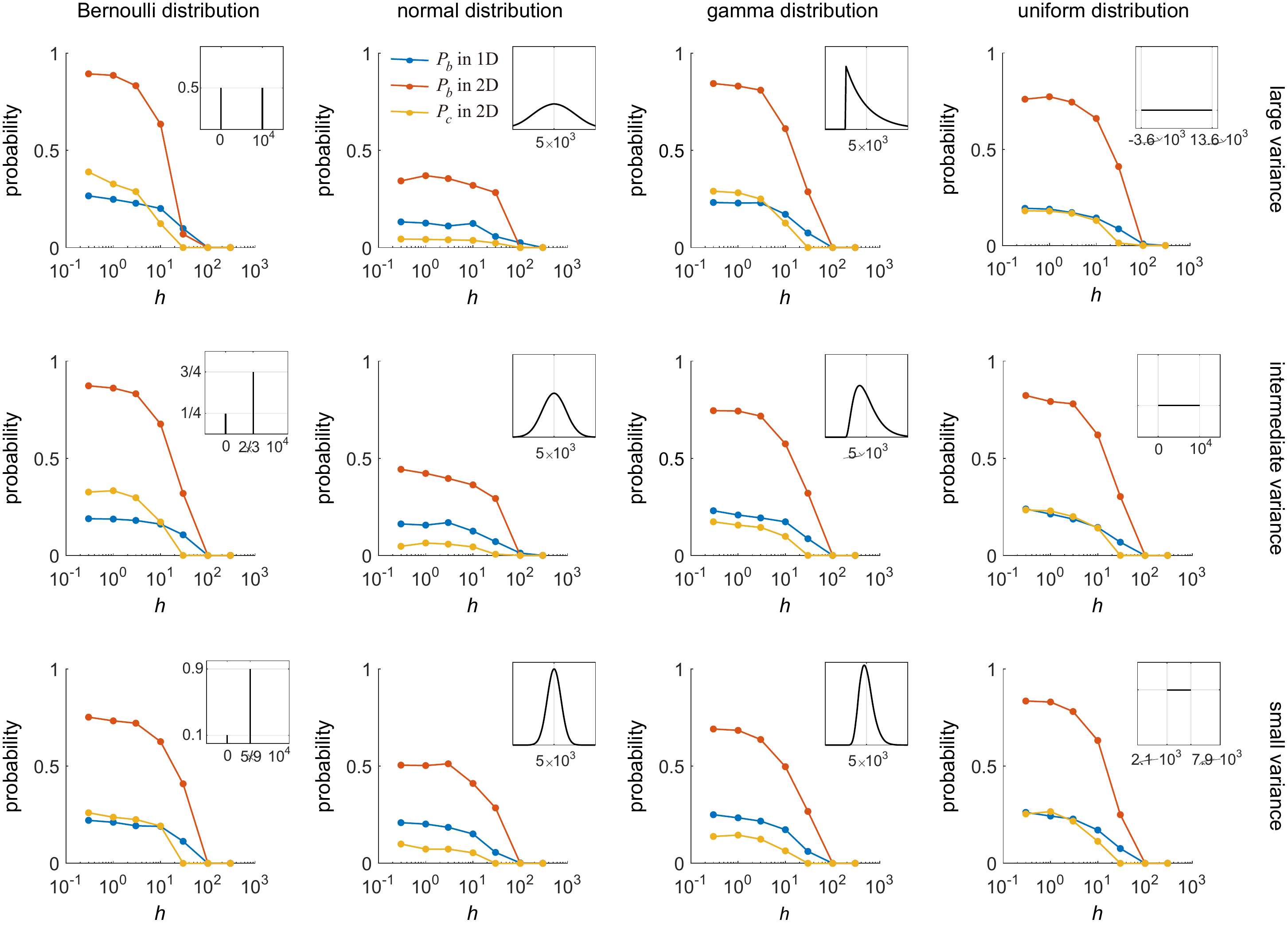}
\caption{\textbf{Boundary effect for four types of random potentials.} The boundary probabilities $P_b$ and $P_c$ as the parameter $h$ varies in the one- and two-dimensional cases for different types of random potentials $V$. The four columns correspond to random potentials sampled from the Bernoulli distribution, the normal distribution, the gamma distribution, and the uniform distribution respectively. The parameter $N$ is chosen as $N = 50$ in the one-dimensional case and $N = 15$ in the two-dimensional case. The insets show the probability mass/density functions of $KV$ (the random potential $V$ multiplied by the strength $K$ of disorder) within each hypercube. The strength of disorder is chosen to be $K = 10^4$, the mean of the random potential is fixed to be $\mu = 1/2$, and the standard deviations of the random potential are chosen to be $\sigma = \mu$, $\sigma = \mu/\sqrt{3}$, and $\sigma = \mu/3$ respectively. The three rows correspond to three different standard deviations.}\label{fig0}
\end{figure}

To answer this, we compare the boundary effects for four types of random potentials sampled from the Bernoulli distribution, the normal distribution, the gamma distribution, and the uniform distribution, respectively, as shown in Fig. \ref{fig0}. Here the strength of disorder is chosen to be $K = 10^4$. For each distribution type, we illustrate the boundary probabilities $P_b$ and $P_c$ in the one- and two-dimensional cases for random potentials with the same mean $\mu = 1/2$ and three different standard deviations, where the standard deviations $\sigma$ within each hypercube are chosen to be $\sigma = \mu$, $\sigma = \mu/\sqrt{3}$, and $\sigma = \mu/3$, respectively (see the four columns of Fig. \ref{fig0}). It is clear that the boundary effect is significant for all types of random potentials. Interestingly, we find that the boundary effect is insensitive to the standard deviation of the random potential within each hypercube. For the Bernoulli and gamma distributions, the boundary probabilities $P_b$ and $P_c$ become slightly higher as the standard deviation decreases, i.e. as the distribution becomes more concentrated. However, the normal and uniform distributions give rise to the opposite effect, i.e. the boundary probabilities become slightly lower with the decrease of the standard deviation. Furthermore, we find that whether the low energy quantum states are localized on the boundary is remarkably affected by the distribution type. The random potentials sampled from the normal distribution lead to a much weaker boundary effect than those sampled from the other three types of distributions.

\section{Multimodality}\label{multimodality}
When $K\gg 1$, we have shown that the original system \eqref{anderson} can be decomposed into many local systems in smaller subregions $D_1, D_2, ..., D_M$. The spectrum of the original system is composed of the eigenvalues of the local systems. If an eigenvalue of the original system coincides with one of the local eigenvalues in the subregion $D_k$, then the associated eigenmode will be localized in $D_k$. Along this line, if multiple subregions $D_{k_1}, D_{k_2}, ..., D_{k_r}$ share a common local eigenvalue, then the corresponding eigenmode may be localized in these subregions simultaneously and thus have multiple peaks. This phenomenon will be referred to as \emph{multimodality} in this paper. Clearly, multimodality takes place when the system has multiple eigenvalues which are approximately equal.

\begin{figure}
\centering
\includegraphics[width=\linewidth]{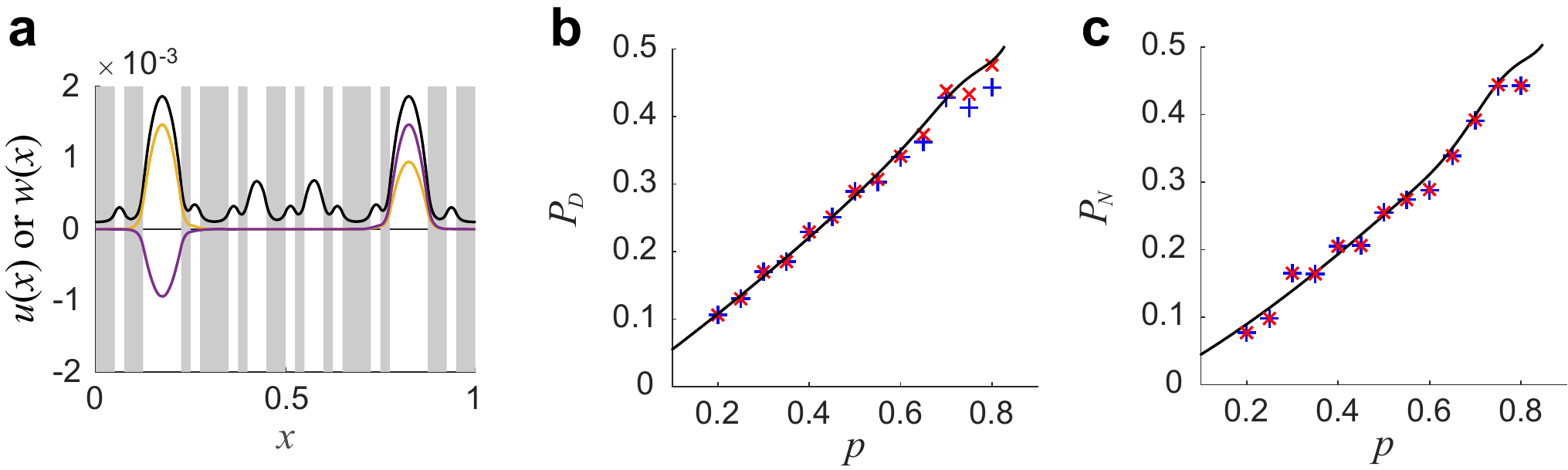}
\caption{\textbf{Multimodality of the eigenmodes in the one-dimensional case}. (a) Multimodality of the first two eigenmodes of a one-dimensional problem under the Neumann boundary condition. The grey parts show the subregions with $V = 1$ and the white parts show the subregions with $V = 0$. The black curve shows the landscape, the yellow curve shows the first eigenmode, and the purple curve shows the second eigenmode. The parameters are chosen as $K = 10^4$ and $N = 50$.
(b) Dependence of the multimodal probability $P_D$ on the parameter $p$ under the Dirichlet boundary condition when $K$ is large. (c) Dependence of the multimodal probability $P_N$ on the parameter $p$ under the Neumann boundary condition when $K$ is large. In (b),(c), the black curve shows the theoretical prediction given in \eqref{multiD} or \eqref{multiN}, the blue plus signs show the frequency at which the first eigenmode displays multimodality, and the red crosses show the frequency at which there are more than one longest extended subregions. Here the frequencies are obtained by generating the random potential 1000 times. The parameters are chosen as $K = 3 \times 10^6$ and $N = 50$.}
\label{fig4}
\end{figure}

Fig. \ref{fig4}(a) illustrates multimodality for a one-dimensional problem. In the one-dimensional case, we have shown that the first eigenmode is localized in the longest extended subregion when $K\gg 1$. If the longest two extended subregions are the same in length, then the first eigenmode may be localized in these two subregions simultaneously. In higher dimensions, any two subregions rarely have the same shape and thus rarely share a common eigenvalue. Hence multimodality in general will not occur in the high-dimensional case. Next we only focus on multimodality in the one-dimensional case.

As mentioned earlier, the domain $\Omega = (0, 1)$ is divided uniformly into $N$ subintervals of the same length $1/N$. The number of subintervals included in each subregion with $V = 0$ ($V = 1$) approximately follows a geometric distribution with parameter $p$ ($q$). When $K\gg 1$, the first eigenmode must be localized in the longest extended subregion with $V = 0$ for both the Dirichlet and Neumann boundary conditions. Thus the probability that the first eigenmode displays multimodality is approximately equal to the probability that there are more than one longest extended subregions. For the Dirichlet boundary condition, the probability of multimodality is approximately given by (see Appendix \ref{appB} for the proof)
\begin{equation}\label{multiD}
P_D \approx 1 - M p \sum_{n=1}^{\infty} (1 - q^{n-1})^{M-1} q^{n-1}.
\end{equation}
For the Neumann boundary condition, the probability of multimodality is much more complicated and is approximately given by
\begin{equation}\label{multiN}
\begin{split}
P_N \approx&\; 1 - q^2 (M-2) \sum_{n=1}^{\infty} (1 - q^{[\frac{n-1}{2}]})^2 (1 - q^{n-1})^{M-3} q^{n-1} p \\
&\;- 2 q^2 \sum_{n=1}^{\infty} (1 - q^{2n-1})^{M-2} (1 - q^{n-1}) q^{n-1} p \\
&\;- 2 p q (M-1) \sum_{n=1}^{\infty} (1 - q^{[\frac{n-1}{2}]}) (1 - q^{n-1})^{M-2} q^{n-1} p \\
&\;- 2 p q \sum_{n=1}^{\infty} (1 - q^{2n-1})^{M-1} q^{n-1} p \\
&\;- p^2 M \sum_{n=1}^{\infty} (1 - q^{n-1})^{M-1} q^{n-1} p,
\end{split}
\end{equation}
where $[x]$ represents the largest integer less than or equal to $x$.


To test our theory, we compare the analytical expressions given in \eqref{multiD} and \eqref{multiN} with numerical simulations for both the Dirichlet and Neumann boundary conditions (Fig. \ref{fig4}(b),(c)).
It can be seen that the theoretical prediction is in excellent agreement with simulation results. For both boundary conditions, the probability of multimodality increases with the parameter $p$ for large $K$. When $p = 0.5$, the probability of multimodality is $0.28$ for the Dirichlet boundary condition and $0.25$ for the Neumann boundary conditions, which means that the first eigenmode has a one in four chance of displaying multimodality. The relatively large values of $P_D$ and $P_N$ also suggest that multimodality is a common phenomenon in the one-dimensional case.

\section{Bifurcation of localization subregions}\label{transition}

\subsection{Bifurcation analysis in a toy model}
From the simulations in Fig. \ref{fig2}(a), we can see that for different values of $K$, the eigenmodes may be localized in completely different subregions. When $K\gg 1$, the first eigenmode is localized in the extended subregion with the smallest local eigenvalue. In particular, in the one-dimensional case, the first eigenmode is localized in the longest extended subregion with $V = 0$. When $K$ is relatively small, however, the first eigenmode may be localized in somewhere else. This suggests that as $K$ increases, the localization subregion for a given eigenmode may change and a \emph{bifurcation phenomenon} may take place. Note that in Fig. \ref{fig2}(a), while the first eigenmode is not localized in the longest extended subregion with $V = 0$ when $K$ is small, it is localized in the union of some long subregions with $V = 0$ that are separated by some short barriers with $V = 1$. The union these subregions is even longer than the longest extended subregion. In other words, when $K$ is large, the localization behavior of the system is very sensitive to the short potential barriers, while it is insensitive to these barriers when $K$ is small. We emphasize that while Fig. \ref{fig2} only shows a bifurcation for the first eigenmode, similar phenomena can be also observed for other eigenmodes.
\begin{figure}[!htb]
\centering
\includegraphics[width=\linewidth]{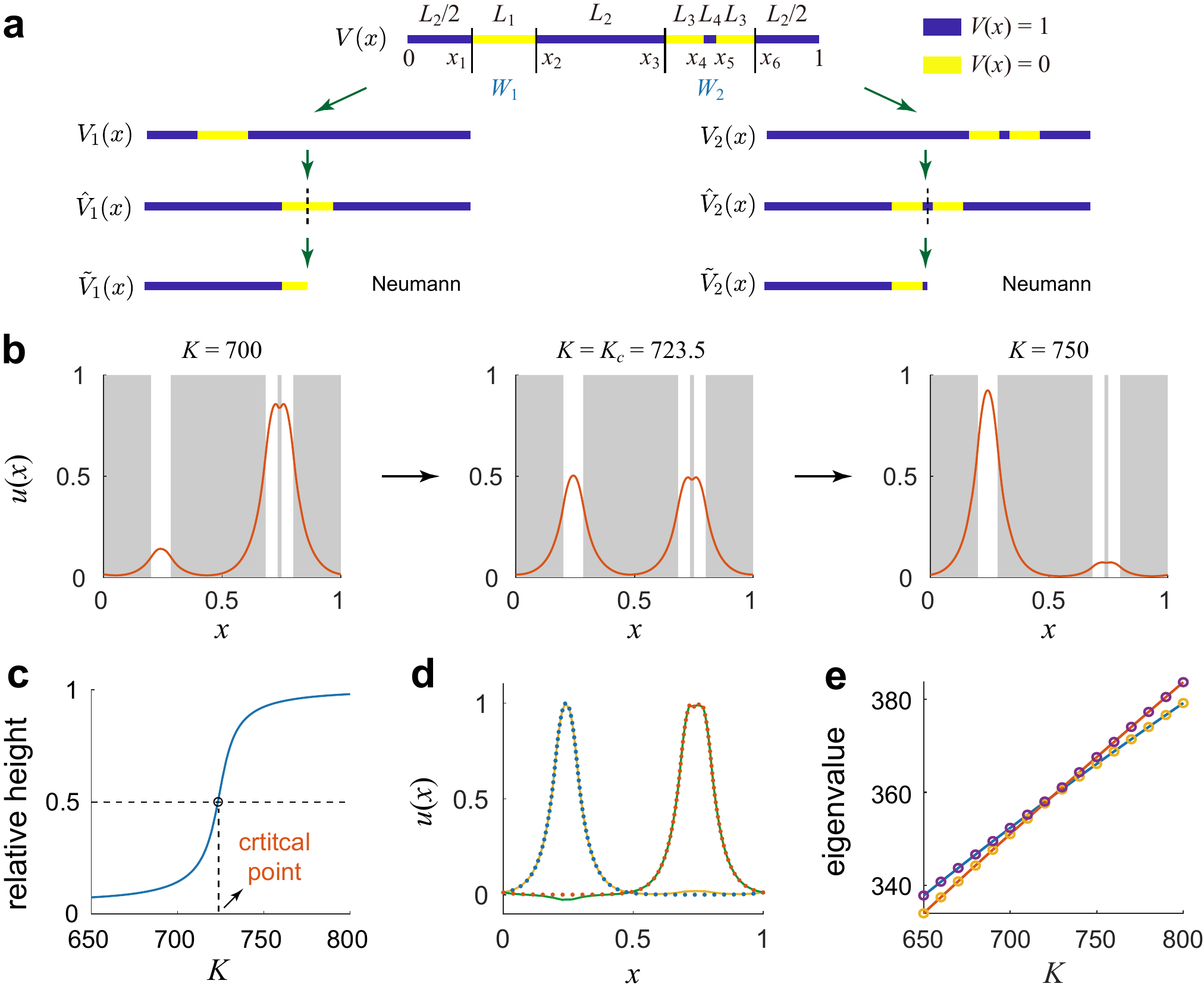}
\caption{\textbf{Bifurcation of localization subregions as $K$ varies}. (a) The potential of the toy model $S$ and the potentials of the subsystems $S_1$ and $S_2$. The yellow parts show the subregions with $V = 1$ and the purple parts show the subregions with $V = 0$. (b) The first eigenmode as $K$ increases. The critical threshold of $K$ is defined as the value at which the two peaks of the eigenmode are equal in height. The eigenmode is bimodal around the critical threshold of $K$. (c) The relative height $F$ of the left peak of the first eigenmode as $K$ varies. (d) The first two eigenmodes (yellow and green curves) of the original system $S$ and the first eigenmodes (blue and red dots) of the two subsystems $S_1$ and $S_2$. Here we choose $K = 800$. (e) The first two eigenvalues (orange and purple circles) of the original system $S$ and the first eigenvalues (blue and red lines) of the two subsystems $S_1$ and $S_2$ under different values of $K$. In (b)-(e), the parameters are chosen as $L_1 = 1/12, L_2 = 2/5, L_3= 1/20, L_4 = 1/60$.}
\label{fig5}
\end{figure}

To gain a deeper insight into the bifurcation phenomenon, we consider a one-dimensional toy model with periodic boundary condition (Fig. \ref{fig5}(a)). In this model, the potential $V$ takes the values of $0$ and $1$ alternately in the intervals with lengths $L_2/2$, $L_1$, $L_2$, $L_3$, $L_4$, $L_3$, and $ L_2/2$ respectively, i.e.
\begin{equation*}
V = \left\{
\begin{split}
1 & \quad x \in [0, x_1), \\
0 & \quad x \in [x_1, x_2), \\
1 & \quad x \in [x_2, x_3), \\
0 & \quad x \in [x_3, x_4), \\
1 & \quad x \in [x_4, x_5), \\
0 & \quad x \in [x_5, x_6), \\
1 & \quad x \in [x_6, 1],
\end{split}
\right.
\qquad
\left\{
\begin{split}
x_1 & = L_2/2, \\
x_2 & = L_2/2 + L_1, \\
x_3 & = L_2/2 + L_1 + L_2, \\
x_4 & = L_2/2 + L_1 + L_2 + L_3, \\
x_5 & = L_2/2 + L_1 + L_2 + L_3 + L_4, \\
x_6 & = L_2/2 + L_1 + L_2 + L_3 + L_4 + L_3.
\end{split}
\right.
\end{equation*}
The periodic boundary condition guarantees that the system has stronger symmetry and thus the theory will become much simpler. The potential $V$ is composed of two ``wells" with lengths $L_1$ and $2L_3+L_4$ respectively, while there is a short barrier with length $L_4$ in the middle of the right well. The distance between the two wells is $L_2$. For convenience, the left and right wells are denoted by $W_1 = [x_1, x_2]$ and $W_2 = [x_3, x_5]$, respectively.

We next investigate the localization of the quantum states of the stationary Schr\"{o}dinger equation
\begin{equation}\label{toy}
S: \;
\left\{
\begin{split}
& -\triangle u + K V u = \lambda u, \quad \textrm{in} \; (0, 1), \\
& \quad u(0) = u(1), \quad u'(0) =  u'(1).
\end{split}
\right.
\end{equation}
Here we require that the left well $W_1$ is the longest subinterval with $V = 0$ and the right well $W_2$ is the union of two relatively long subintervals with $V = 0$ that are separated by a very short subinterval with $V = 1$. The total length of the two relatively long subintervals should be greater than the length of the longest subinterval so that bifurcation can occur. To ensure that the two wells do not interfere each other, we further require the distance between the two wells to be large enough. To be more specific, the parameters $L_1$, $L_2$, $L_3$, and $L_4$ should satisfy: (i) $L_1 > L_3$, which grantees that $W_1$ is the longest subinterval; (ii) $L_1 < 2L_3$, which grantees that $W_2$ is longer than $W_1$; (iii) $L_4 < L_3/2$, which guarantees that the barrier in $W_2$ is short enough; (iv) $L_2 > L_1 + 2L_3 + L_4$, which grantees that the two wells are far enough; (v) $L_1 + 2L_2 + 2 L_3 + L_4 = 1$, which guarantees that the length of the whole interval is $1$.

For the toy model, we illustrate the first eigenmode of the system $S$ given in \eqref{toy} under different values of $K$ (Fig. \ref{fig5}(b)). Since the right well is longer than the left one and the barrier in the right well is short, when $K$ is small, the first eigenmode is mainly localized in the right well. Since the left well is the longest subinterval with $V = 0$, when $K$ is large, the first eigenmode is mainly localized in the left well. With the increase of $K$, the left peak of the eigenmode becomes higher and the right peak becomes lower. At a critical threshold of $K$, the two peaks are equal in height and bifurcation takes place --- the confining subregion of the eigenmode transitions from the right well to the left one. For the first eigenmode $u$, we introduce
\begin{equation}
F = \frac{\max_{x \in W_1} |u(x)|}{\max_{x \in W_1} |u(x)| + \max_{x \in W_2} |u(x)|},
\end{equation}
which represents the height of the left peak relative to the total height of the two peaks. In general, the relative height $F$ of the left peak for the first eigenmode increases with $K$ (Fig. \ref{fig5}(c)). The critical threshold of $K$ is defined as the value at which $F = 0.5$. At the critical value $K = K_c$, the relative height $F$ of the left peak changes sharply from a low to a high value, corresponding to the occurrence of bifurcation.

When $L_2$ is large, the two wells are far apart and they have little influence on each other. Hence we can decompose the system $S$ into two subsystems $S_1$ and $S_2$ with $S_1$ corresponding to the left peak of the eigenmode and with $S_2$ corresponding to the right peak. Specifically, we introduce two new potentials $V_1$ and $V_2$, where $V_1$ is obtained from $V$ by setting the values in the right well to $1$ and $V_2$ is obtained from $V$ by setting the values in the left well to $1$ (Fig. \ref{fig5}(a)), i.e.
\begin{equation*}
V_1(x) = \left\{
\begin{split}
1 & \quad x \in [0, x_1), \\
0 & \quad x \in [x_1, x_2), \\
1 & \quad x \in [x_2, 1],
\end{split}
\right.
\qquad
V_2(x) = \left\{
\begin{split}
1 & \quad x \in [0, x_3), \\
0 & \quad x \in [x_3, x_4), \\
1 & \quad x \in [x_4, x_5), \\
0 & \quad x \in [x_5, x_6), \\
1 & \quad x \in [x_6, 1].
\end{split}
\right.
\end{equation*}
Then the original system $S$ can be decomposed into two subsystems $S_1$ and $S_2$, which are defined as
\begin{equation}
S_i: \;
\left\{
\begin{split}
& -\triangle u + K V_i u = \lambda u, \quad \textrm{in} \; (0, 1), \\
& \quad u(0) = u(1), \; u'(0) =  u'(1),
\end{split}
\right.
\quad
i = 1, 2.
\end{equation}
Here the subsystem $S_1$ has potential $V_1$ and the subsystem $S_2$ has potential $V_2$. Intuitively, for the first two eigenmodes of the system $S$, one is localized in the left well $W_1$ and the other is localized in the right well $W_2$. The eigenmode localized in $W_1$ ($W_2$) can be approximated by the first eigenmode of the subsystem $S_1$ ($S_2$). Fig. \ref{fig5}(d) illustrates the first two eigenmodes of the original system and the first eigenmodes of the two subsystems. It can be seen that the first eigenmode of each subsystem almost coincides with one of the first two eigenmodes of the original system. Moreover, we illustrate the first two eigenvalues of the original system and the first eigenvalues of the two subsystems under different values of $K$ in Fig. \ref{fig5}(e). Clearly, the first eigenvalues of the two subsystems serve as good approximations to the first two eigenvalues of the original system.

We now provide a detailed explanation for the occurrence of bifurcation. From Fig. \ref{fig5}(e), the first eigenvalue of each subsystem increases with $K$. However, the eigenvalues of both subsystems increase with $K$ at different rates (see two lines of different slopes in Fig. \ref{fig5}(e)). When $K$ is small, the eigenvalue of the subsystem $S_2$ is smaller than that of the subsystem $S_1$, and thus the first eigenmode of the original system $S$ is localized in the right well $W_2$. When $K$ is large, the eigenvalue of $S_2$ exceeds that of $S_1$, and thus thus the first eigenmode of $S$ is localized in the left well $W_1$. At the critical threshold $K = K_c$, the localization subregion of the first eigenmode transitions from $W_2$ to $W_1$. In analogy to multimodality, bifurcation occurs when the first eigenvalues of both subsystems are approximately equal, i.e. when the first and the second eigenvalues of the original system are approximately equal. This critical eigenvalue is denoted by $\lambda = \lambda_c$.

\subsection{Computation of the critical threshold}
Clearly, the critical threshold of $K$ plays a crucial role in the bifurcation phenomenon. Numerically, in order to compute the critical value, we need to solve the original system under many different values of $K$, which is very time-consuming. Here we provide an analytical theory of the critical threshold and the associated critical eigenvalue.

To this end, we first simplify the subsystems $S_1$ and $S_2$. For each subsystem $S_i$, due to the periodic boundary condition, we can shift the potential $V_i$ along the interval $\Omega = (0,1)$ without changing the eigenvalues. In particular, we move the well $W_i$ to the middle of the interval (Fig. \ref{fig5}(a)) and the shifted potentials for the two subsystems are defined as
\begin{equation*}
\hat{V}_1 = \left\{
\begin{split}
1 & \quad x \in \left[ 0, \frac{1 - L_1}{2} \right), \\
0 & \quad x \in \left[ \frac{1 - L_1}{2}, \frac{1 + L_1}{2} \right), \\
1 & \quad x \in \left[ \frac{1 + L_1}{2}, 1 \right].
\end{split}
\right.
\quad
\hat{V}_2 = \left\{
\begin{split}
1 & \quad x \in \left[ 0, \frac{1 - L_4 - 2 L_3}{2} \right), \\
0 & \quad x \in \left[ \frac{1 - L_4 - 2 L_3}{2}, \frac{1 - L_4}{2} \right), \\
1 & \quad x \in \left[ \frac{1 - L_4}{2}, \frac{1 + L_4}{2} \right), \\
0 & \quad x \in \left[ \frac{1 + L_4}{2}, \frac{1 + L_4 + 2 L_3}{2} \right), \\
1 & \quad x \in \left[ \frac{1 + L_4 + 2 L_3}{2}, 1 \right].
\end{split}
\right.
\end{equation*}
The subsystem with the shifted potential $\hat{V}_i$ is denoted by $\hat{S}_i$, which can be written explicitly as
\begin{equation*}
\hat{S}_i: \;
\left\{
\begin{split}
& -\triangle u + K\hat{V}_i u = \lambda u, \quad \textrm{in} \; (0, 1), \\
& \quad u(0) = u(1), \; u'(0) =  u'(1),
\end{split}
\right.
\quad
i = 1, 2.
\end{equation*}
Note that the shifted potential $\hat{V}_i$ satisfies $\hat{V}_i(x) = \hat{V}_i(1-x)$. Thus the eigenmodes of the subsystem $\hat{S}_i$ satisfy $u(x) = u(1-x)$ and $u'(x) = -u'(1-x)$. Taking $x = 1/2$, we obtain $u'(1/2) = 0$. Moreover, taking $x = 0$ and using the periodic boundary condition $u'(0) = u'(1)$, we obtain $u'(0) = 0$. Then the eigenvalues of the subsystem $\hat{S}_i$ coincide with the eigenvalues of the following equivalent subsystem in the interval $(0, 1/2)$ with the Neumann boundary condition:
\begin{equation}\label{tildeS}
\tilde{S}_i: \;
\left\{
\begin{split}
& -\triangle u + K\tilde{V}_i u = \lambda u, \quad \textrm{in} \; (0, 1/2), \\
& \quad u'(0) = u'(1/2) =  0,
\end{split}
\right.
\quad
i = 1, 2,
\end{equation}
where the potential $\tilde{V}_i$ of the equivalent subsystem is defined as (see Fig. \ref{fig5}(a) for an illustration)
\begin{equation*}
\tilde{V}_i(x) = \hat{V}_i(x), \quad x \in (0, 1/2), \; i = 1, 2.
\end{equation*}

After the above simplification, the eigenvalues of each subsystems $S_i$ are consistent with those of the equivalent subsystem $\tilde{S}_i$. For a fixed value of $K$, each eigenvalue $\lambda$ of the subsystem $\tilde{S}_1$ satisfies the equation (see Appendix \ref{appC} for the proof)
\begin{equation*}
D_1(K, \lambda) = \alpha \tan(\alpha t_0) - \beta \tanh(\beta (1/2 - t_0)) = 0,
\end{equation*}
and each eigenvalue $\lambda$ of the subsystem $\tilde{S}_2$ satisfies the equation
\begin{equation*}
D_2(K, \lambda) = (\alpha^2 - \beta^2)\frac{\mathrm{e}^{2 \beta t_2} + \mathrm{e}^{2 \beta (t_1+t_3)}}
{\mathrm{e}^{2 \beta (t_1+t_3)} - \mathrm{e}^{2 \beta t_2}}
+ (\alpha^2 + \beta^2)\frac{\mathrm{e}^{2 \beta t_3} + \mathrm{e}^{2 \beta (t_1+t_2)}}{\mathrm{e}^{2 \beta (t_1+t_3)} - \mathrm{e}^{2 \beta t_2}} + 2 \alpha \beta \cot(\alpha (t_1 - t_2)) = 0,
\end{equation*}
where $\alpha = \sqrt{\lambda}$, $\beta = \sqrt{K - \lambda}$, $t_0 = L_1 / 2$, $t_1 = L_4 / 2$, $t_2 = L_4 / 2 + L_3$, and $t_3 = 1 / 2$. At the critical threshold $K = K_c$, the first eigenvalues of the two subsystems are approximately equal. Then the critical threshold $K_c$ and the critical eigenvalue $\lambda_c$ approximately satisfy the following system of algebraic equations:
\begin{equation}\label{phase0}
\left\{
\begin{split}
D_1(K_c, \lambda_c) = 0, \\
D_2(K_c, \lambda_c) = 0.
\end{split}
\right.
\end{equation}
This system of equations can be used to analytically predict the critical threshold $K_c$ and the critical eigenvalue $\lambda$ of bifurcation.

To test our theory, we sample the parameters $L_1$, $L_2$, $L_3$, and $L_4$ randomly from a large swathe of the parameter space. Specifically, the parameters are chosen as $L_3 \sim U[0.03, 0.055]$, $L_1 \sim U[1.3 L_3, 1.7 L_3]$, $L_4 \sim U[0.2 L_3, 0.4 L_3]$, and $L_2$ is determined so that $L_1 + 2 L_2 + 2 L_3 + L_4 = 1$, where $U[a,b]$ denotes the uniform distribution over the interval $[a,b]$. We then use \eqref{phase0} to predict $K_c$ and $\lambda_c$. For $1000$ sample points, the mean relative prediction errors ((predicted value $-$ real value) $/$ real value) for $K_c$ and $\lambda_c$ are only $1.52 \times 10^{-4}$ and $1.61 \times 10^{-4}$, respectively. This shows that our analytical theory can indeed capture the critical values of bifurcation accurately without carrying out numerical simulations.

\subsection{Dependence of the critical threshold on model parameters}
We have seen that the critical threshold $K_c$ and critical eigenvalue $\lambda_c$ satisfy the system of algebraic equations \eqref{phase0}. Clearly, the value of $K_c$ depends on the parameters $L_1$, $L_2$, $L_3$, and $L_4$. However, the relationship between them is very complicated. Here we investigate their relationship numerically and reveal the key factors of bifurcation.

\begin{figure}
\centering
\includegraphics[width=\linewidth]{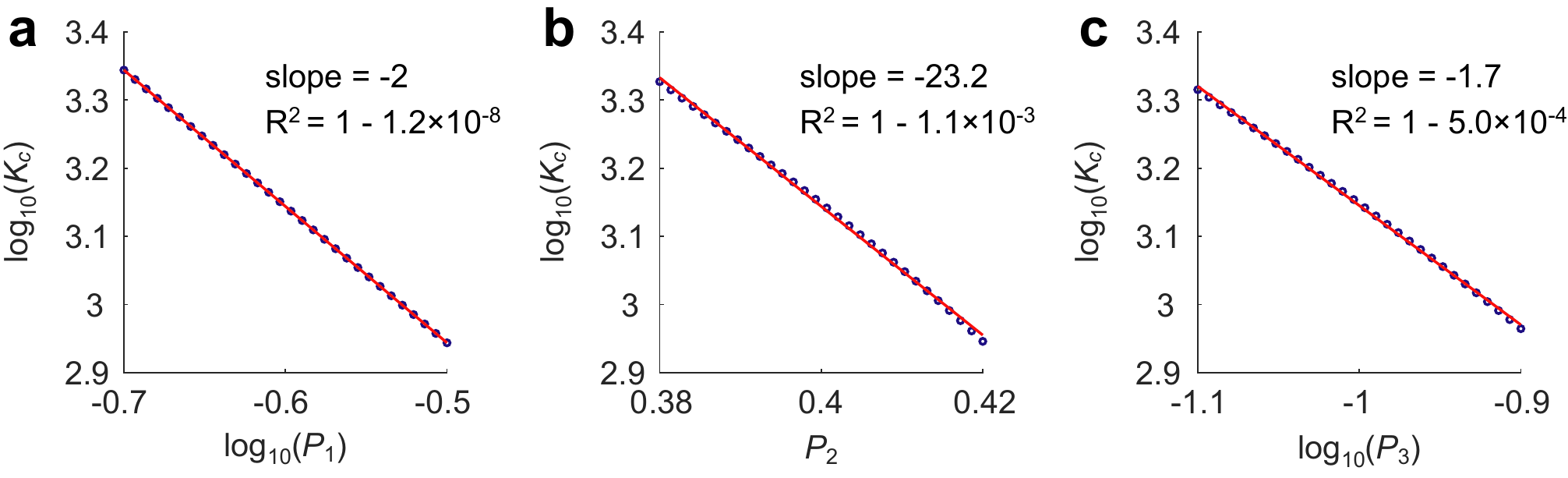}
\caption{\textbf{Influence of model parameters on the critical threshold $K_c$}. (a) Dependence of $K_c$ on $P_1$. Here 30 points are chosen uniformly from the interval $[-0.7,-0.5]$ for the value of $\log_{10}(P_1)$. (b) Dependence of $K_c$ on $P_2$. Here 30 points are chosen uniformly from the interval $[0.38,0.42]$ for the value of $P_2$. (c) Dependence of $K_c$ on $P_3$. Here 30 points are chosen uniformly from the interval $[-1.1,-0.9]$ for the value of $\log_{10}(P_3)$. In (a)-(c), the blue circles show the simulation results and the red lines show the corresponding regression line. Except the one that is tuned, the other parameters are chosen as $P_1 = 0.25, P_2 = 0.4, P_3 = 0.1$.}
\label{fig6}
\end{figure}

Since $L_1 + 2 L_2 + 2 L_3 + L_4 = 1$, there four parameters only have three degrees of freedom. For simplicity, we define
\begin{equation}
P_1 = \frac{L_1 + 2 L_3 + L_4}{L_1 + 2 L_2 + 2 L_3 + L_4}, \quad P_2 = \frac{L_1}{L_1 + 2 L_3 + L_4}, \quad P_3 = \frac{L_4}{2 L_3 + L_4},
\end{equation}
where $P_1$ represents the total length of the two wells relative to the length of the whole interval, $P_2$ represents the length of the left well relative to total length of the two wells, and $P_3$ represents the length of the barrier relative to the length of the right well. The relationship between $K_c$ and $P_1$, $P_2$, and $P_3$ is complicated. However, we can obtain an approximate relationship between them using our analytical theory. Fig. \ref{fig6}(a)-(c) illustrate the critical threshold $K_c$ as $P_1$, $P_2$, and $P_3$ vary. Here we generate different values of $P_1$, $P_2$, or $P_3$ randomly from a large swathe of the parameter space and keep the other two parameters invariant. Interestingly, we find that $K_c$ depends on $P_1$ in a power law form as
\begin{equation}
K_c \propto {P_1}^{-2}.
\end{equation}
To check this, we plot $\log(K_c)$ as a function of $\log(P_1)$ in Fig. \ref{fig6}(a) and find that there is an excellent linear relationship between them with an exceptionally high $R^2$ and a slope of $-2$. Similarly, we find that $K_c$ depends on $P_2$ in an exponential form as (Fig. \ref{fig6}(b))
\begin{equation}
K_c \propto e^{-23.2P_2},
\end{equation}
and $K_c$ depends on $P_3$ in a power law form as (Fig. \ref{fig6}(c))
\begin{equation}
K_c \propto {P_3}^{-1.7}.
\end{equation}
These results indicate that $K_c$ decreases with $P_1$ and $P_3$ at a power law speed and decreases with $P_2$ at an exponential speed. In particular, short potential wells (small $P_1$), a short left well relative to the right well (small $P_2$), and a short barrier in the right well (small $P_3$) are more capable of producing a large critical value of $K$. The exponential decay of $K_c$ with $P_2$ suggests that the relative length of the left well affects the critical threshold more severely than the other two factors.

\section{Conclusions and discussion}\label{conclusion}
In the present paper, we focused on some new mathematical aspects of Anderson localization for the quantum states of a Schr\"{o}dinger equation with a disordered lattice potential (or more generally, the eigenmodes of a second-order elliptic operator with a random potential). In previous papers \cite{filoche2012universal}, the concepts of localization landscape and valley lines have been proposed  for symmetric elliptic operators and the Dirichlet boundary condition. Here we generalized these concepts to non-symmetric elliptic operators and general boundary conditions using the probabilistic representation of the solution of a second-order elliptic equation by reflecting diffusions. We showed that the confining subregions of low energy quantum states are perfectly predicted by the peak positions of the landscape and are enclosed by the valley lines of the landscape.

We also studied the localization behavior of the quantum states when the strength $K$ of disorder is large. In the limit of $K\rightarrow\infty$, we demonstrated that the eigenmodes must be localized in the region with the lowest potential. When the potential is sampled from the Bernoulli distribution, this region can be decomposed into many connected components (subregions), which are exactly separated by the valley lines of the landscape. Then the original system can be decomposed into many uncoupled local systems in these subregions. We further showed that the spectrum of the original system is composed of the eigenvalues of the local systems. If an eigenvalue of the original system coincides with one of the local eigenvalue in a particular subregion, then the corresponding eigenmode will be localized in that subregion. In particular, if multiple subregions share a common local eigenvalue, then the corresponding eigenmode will have multiple peaks and thus display multimodality. In the one-dimensional case, we derived the analytical expression for the probability that the first eigenmode displays multimodality when $K$ is large under the Dirichlet or Neumann boundary condition, which was then used to study the influence of the parameter $p$ on multimodality.

For Neumann and Robin boundary conditions, we found that (i) compared to the interior of the domain, the eigenmodes tend be localized on the boundary; (ii) compared to the one-dimensional case, the eigenmodes for a two-dimensional problem are more likely be localized on the boundary, especially when $h$ is small and $K$ is large. To explain this phenomenon, we proposed the concept of extended subregions, which are obtained from the original ones through mirror symmetry along the boundary, and showed that the low energy eigenmodes are more likely to be localized in large extended subregions with strong symmetry. Since (i) the extended subregion should in general be doubled when the original subregion is located on the boundary and (ii) the extended subregion in general has strong symmetric than the original ones, the eigenmodes tend be localized on the boundary. Moreover, since (i) the proportion of subregions that need to extend are much higher in the two-dimensional case and (ii) the extended subregion should be magnified four times when the original subregion is located in the corner, the eigenmodes for a two-dimensional problem are more likely to be localized on the boundary. In the one-dimensional case, we also derived an explicit expression for the probability that the first eigenmode is localized on the boundary, which was then used to examine the influence of the parameter $p$ on the boundary effect.

When $K$ is large, the first eigenmode is localized in the extended subregion with the smallest local eigenvalue. In particular, in the one-dimensional case, the first eigenmode is localized in the longest extended subregion with $V = 0$ for both the Dirichlet and Neumann boundary conditions. However, when $K$ is small, the eigenmode is often localized in another region that is the union of some subregions with $V = 0$ that are separated by some short barriers with $V = 1$. At the critical value of $K$, the localization behavior of the system changes severely and bifurcation occurs. To gain a deeper insight into bifurcation, we consider a one-dimensional toy model whose potential has two wells: the left one is the longest subinterval with $V = 0$ and the right one is the union of two relative long subintervals with $V = 0$ separated by a short barrier with $V = 1$ in between. For the toy model, we showed that bifurcation occurs when its first and second eigenvalues are approximately equal. Based on this property, we further obtained the equations satisfied by the critical threshold $K = K_c$ and the critical eigenvalue $\lambda = \lambda_c$. The analytical theory is then used to study the dependence of bifurcation on model parameters and reveal the key factors of bifurcation.

Due to the complexity of Anderson localization, we mainly develop the analytical theory of the boundary effect, multimodality, and bifurcation in the one-dimensional case when the lattice potential is randomly sampled from a Bernoulli distribution. In the study of bifurcation, we consider only a toy model whose potential has two wells with the right well having a short barrier in the middle. We anticipate that our results can be generalized to high dimensions and more general random potentials, and further stimulate work in this area. We also hope that the landscape theory and the probabilistic approach can be applied to high-frequency localization behavior \cite{grebenkov2013geometrical, jia2021two} and eigenvalue problems in other disciplines such as biology and chemistry \cite{jia2014overshoot, jia2016nonequilibrium, jia2018relaxation}.

\section*{Acknowledgements}
The authors acknowledge the support from National Natural Science Foundation of China with
grants No. 11871092, No. 12131005, and NSAF U1930402.

\begin{appendices}

\section{Proof of the Filoche-Mayboroda inequality}\label{appA}
Let $(X, F) = (X_t,F_t)_{t\geq 0}$ be the solution to the Skorokhod stochastic differential equation
\begin{equation*}
d X_t = b(X_t) dt + \sigma(X_t) d W_t - g n(X_t) d F_t,
\end{equation*}
where $a = \sigma\sigma^T$ is the diffusion matrix and $F_t$ is a continuous non-decreasing process that increases only when $X_t \in \partial \Omega$. For convenience, we define
\begin{equation*}
Y_t = e^{-\int_{0}^{t} h(X_s) d F_s - \int_{0}^{t} K V(X_s) ds}.
\end{equation*}
By Ito's formula \cite{protter2005stochastic}, we have
\begin{equation*}
\begin{split}
d[u(X_t) Y_t] = &\; \partial_i u(X_t) Y_t dX^i_t + \frac{1}{2} \partial_{ij} u(X_t) Y_t (dX^i_t) (dX^j_t) - u(X_t) Y_t [h(X_t) dF_t + K V(X_t) dt] \\
= &\; \partial_i u(X_t) b^i(X_t) Y_t dt + \partial_i u(X_t) \sigma^i_j(X_t) Y_t dW^j_t - \partial_i u(X_t) g n^i(X_t) Y_t dF_t\\
&\; + \frac{1}{2} \partial_{ij} u(X_t) a^{ij}(X_t) Y_t dt - u(X_t) Y_t [h(X_t) dF_t + K V(X_t) dt]\\
= &\; \partial_i u(X_t) \sigma^i_j(X_t) Y_t dW^j_t + (L u - K V u) (X_t) Y_t dt
- \left( g \frac{\partial u}{\partial n} + h u \right) (X_t) Y_t dF_t,
\end{split}
\end{equation*}
where we have used Einstein's summation convention: if the same index appears twice in any term, once as an upper index and once as a lower index, that term is understood to be summed over all possible values of that index. Since $F_t$ only increases when $X_t \in \partial \Omega$ and since $g \partial u/\partial n + h u = 0$ on $\partial \Omega$, we obtain
\begin{equation*}
d[u(X_t) Y_t] = \partial_i u(X_t) \sigma^i_j(X_t) Y_t dW^j_t - \lambda u(X_t) Y_tdt.
\end{equation*}
This shows that
\begin{equation*}
u(X_t)Y_t = u(X_0) + \int_0^t \partial_i u(X_s) \sigma^i_j(X_s) Y_s dW^j_s - \lambda \int_0^t u(X_s) Y_s ds.
\end{equation*}
Note that the middle term on the right-hand side of the above equation is a martingale. Thus we have
\begin{equation*}
u(x) = \Enum_xu(X_t)Y_t+\lambda\Enum_x\int_0^t u(X_s)Y_sds.
\end{equation*}
Taking $t\rightarrow\infty$ in the above equation yields
\begin{equation}\label{temp1}
u(x) = \lambda\Enum_x\int_0^\infty u(X_s)Y_sds.
\end{equation}
Recall that the localization landscape $w = w(x)$ is defined as the solution to the following PDE:
\begin{equation*}
\left\{
\begin{split}
& - L w + K V w = 1 \quad \textrm{in} \; \Omega, \\
& \quad g \frac{\partial w}{\partial n} + h w = 0 \quad \textrm{on} \; \partial \Omega,
\end{split}\right.
\end{equation*}
Similarly, the landscape has the following probabilistic representation:
\begin{equation}\label{temp2}
w(x) = \Enum_x\int_0^\infty Y_sds.
\end{equation}
Comparing \eqref{temp1} and \eqref{temp2}, we obtain the Filoche-Mayboroda inequality.

\section{Equivalent system in the extended subregion}\label{appD}
Let $D \subset \Omega$ be an open subset of $\Rnum^n$, where
\begin{equation*}
\Omega = \{x = (x_1, x_2, \cdots, x_n) \in \Rnum^n \, : \, x_1, x_2, \cdots, x_d > 0\}, \quad 0 < d \leq n,
\end{equation*}
is the first quadrant of a $d$-dimensional hyperplane of $\Rnum^n$. Then the extended subregion $D^{(e)}$ of $D$ is defined as the region obtained from $D$ through mirror symmetry with respect to the corresponding hyperplane, i.e.
\begin{equation*}
D^{(e)} = \mathrm{int}(\overline{D^{(e)}_0}), \quad D^{(e)}_0 = \{x \in \Rnum^n \, : \, \tau x \in D \},
\end{equation*}
where $\mathrm{int}(A)$ denotes the interior of the set $A$ and the operator $\tau$ is defined as
\begin{equation*}
\tau (x_1, x_2, \cdots, x_n) = (|x_1|, |x_2|, \cdots, |x_d|, x_{d+1}, \cdots, x_n).
\end{equation*}
We next consider two eigenvalue problems in $D$ and $D^{(e)}$, respectively. The eigenvalue problem in $D$ is given by
\begin{equation}\label{eigenD}
\left\{
\begin{split}
& -\triangle u = \lambda u, \quad \textrm{in} \; D, \\
& \quad \frac{\partial u}{\partial n} = 0, \quad \textrm{on} \; \partial D \cap \partial \Omega,\\
& \quad u = 0, \quad \textrm{on} \; \partial D \setminus \partial \Omega.
\end{split}
\right.
\end{equation}
The eigenvalue problem in $D^{(e)}$ is given by
\begin{equation}\label{eigenDe}
\left\{
\begin{split}
& -\triangle u = \lambda u, \quad \textrm{in} \; D^{(e)}, \\
& \quad u = 0, \quad \textrm{on} \; \partial D^{(e)}.
\end{split}
\right.
\end{equation}
Let $u = u(x)$ be the first eigenmode of the problem \eqref{eigenD} corresponding to the eigenvalue $\lambda$. Then we can define the following function $u^{(e)}$ from $u$ through mirror symmetry:
\begin{gather*}
u^{(e)}(x) = u(x), \quad x \in D,\\
u^{(e)}(x) = u(\tau x), \quad x \in D^{(e)} \setminus D.
\end{gather*}
Clearly, $u^{(e)}$ is sufficiently smooth in $D^{(e)}$. For each $x \in D$, we have
\begin{equation*}
-\triangle u^{(e)}(x) = -\triangle u(x) = \lambda u(x) = \lambda u^{(e)}(x),
\end{equation*}
and for each $x \in D^{(e)} \setminus D$, we have
\begin{equation*}
-\triangle u^{(e)}(x) = -\triangle u(\tau x) = -\triangle u(x) = \lambda u(x) = \lambda u(\tau x) = \lambda u^{(e)}(x).
\end{equation*}
Moreover, it is easy to see that $u^{(e)}$ vanishes on the boundary of $D^{(e)}$. This shows that $u^{(e)}$ is an eigenmode of the problem \eqref{eigenDe} corresponding to the eigenvalue $\lambda$. According to the results in \cite{grebenkov2013geometrical}, the first eigenmode of a Laplacian eigenvalue problem does not change its sign in the domain. Therefore, the eigenmode $u$ can be chosen so that $u(x) > 0$ and thus we have $u^{(e)}(x) > 0$. Due to orthogonality of the eigenmodes, the first eigenmode is the only eigenmode that does not change its sign. Then we have proved that $u^{(e)}(x)$ is the first eigenmode of the problem \eqref{eigenDe}.

\section{Boundary effect and multimodality}\label{appB}
In the one-dimensional case, the domain $\Omega = (0, 1)$ is divided uniformly into $N$ subintervals of the same length $1/N$. In each subinterval, the potential $V$ is sampled from the Bernoulli distribution with parameter $p$. The values of $V$ in these subintervals are independent of each other. Suppose that the values of $V$ in the previous two subintervals are $1$ and $0$, respectively. Let $X$ denote the number of the successive subintervals in which $V$ takes the value of $0$. Clearly, the event of $X = n$ means that $V = 0$ in the following $n-1$ subintervals and $V = 1$ in the $n$th subinterval. Due to the independence of $V$ in these subintervals, when $N$ is sufficiently large, the random variable $X$ approximately follows the geometric distribution with parameter $p$, i.e.
\begin{equation*}
\mathbb{P}(X = n) = q^{n-1} p,
\end{equation*}
where $q = 1 - p$. Similarly, let $Y$ denote the number of successive subintervals in which $V$ takes the value $1$. Then when $N$ is sufficiently large, the random variable $X$ approximately follows the geometric distribution with parameter $q$, i.e.
\begin{equation*}
\mathbb{P}(Y = n) = p^{n-1} q.
\end{equation*}
In this way, the domain $\Omega = (0,1)$ can be decomposed into many subregions with $V = 0$ and $V = 1$ alternatively. Two successive subregions are collectively called a \emph{period}. Clearly, the mean length of each period is given by
\begin{equation*}
\frac{1}{N}\Enum(X+Y) = \frac{1}{Npq}.
\end{equation*}
Therefore, the mean number of periods in the domain is roughly given by $M = N p q$. Clearly, the domain is composed of subregions with $V = 0$ and $V = 1$ alternatively. Let $X_1, X_2, ..., X_M$ denote the numbers of subintervals included in the successive subregions with $V = 0$. When $N$ is sufficiently large, these random variables are approximately independent.

\subsection{Calculation of boundary probability for the Neumann boundary condition}\label{appB1}
Under the Neumann boundary condition, the first eigenmode is localized in the longest extended subregion with $V = 0$ when $K$ is large. Therefore, the event that the first eigenmode is localized on the boundary is equivalent to the event that the longest extended subregion with $V = 0$ appears on the boundary. This event can be classified into the following three situations.

First, we consider the situation where $V(0) = V(1) = 0$, whose probability is $q^2$. In this case, the lengths of the extended subregions with $V = 0$ are given by $2 X_1, X_2, \cdots, X_{M-1}, 2 X_M$. Then the probability that the longest extended subregion appears on the boundary can be computed explicitly as
\begin{equation*}
\begin{split}
p_1 &= \mathbb{P}(\max\{X_2, X_3, \cdots, X_{M-1}\} < 2 \max\{X_1, X_M\}) \\
&= \sum_{m,n=1}^{\infty} \mathbb{P}(\max\{X_2, X_3, \cdots, X_{M-1}\} < 2 \max\{m,  n\}) \mathbb{P}(X_1 = m) \mathbb{P}(X_M = n) \\
&= \sum_{m,n=1}^{\infty} [\mathbb{P}(X_k < 2 \max\{m,n\}) ]^{M-2} \mathbb{P}(X_1 = m) \mathbb{P}(X_M = n)\\
&= \sum_{m,n=1}^{\infty} (1 - q^{2 \max\{m,n\}-1})^{M-2} q^{m-1} p q^{n-1} p\\
&= p^2\sum_{k=1}^{\infty} q^{k-2} \sum_{n=1}^{k-1} (1 - q^{2 \max\{k-n,n\}-1})^{M-2}.
\end{split}
\end{equation*}
Second, we consider the situation where $V(0) = 0$ and $V(1) = 1$, whose probability is $p q$. In this case, the lengths of the extended subregions with $V = 0$ are given by $2 X_1, X_2, \cdots, X_M$. Then the probability that the longest extended subregion appears on the boundary can be computed explicitly as
\begin{equation*}
\begin{split}
p_2 &= \mathbb{P}(\max\{X_2, X_3, \cdots, X_{M}\} < 2 X_1) \\
&= \sum_{n=1}^{\infty} \mathbb{P}(\max\{X_2, X_3, \cdots, X_{M-1}\} < 2 n) \mathbb{P}(X_1 = n) \\
&= \sum_{n=1}^{\infty} [\mathbb{P}(X_k < 2 n)]^{M-1} \mathbb{P}(X_1 = n)\\
&= p \sum_{n=1}^{\infty} (1-q^{2n-1})^{M-1} q^{n-1}.
\end{split}
\end{equation*}
Third, we consider the situation where $V(0) = 1$ and $V(1) = 0$, whose probability is $p q$. In this case, the lengths of the extended subregions with $V = 0$ are given by $X_1, X_2, \cdots, 2X_M$. In analogy to the second case, the probability that the longest extended subregion appears on the boundary is given by
\begin{equation*}
p_3 = \mathbb{P}(\max\{X_1, X_2, \cdots, X_{M-1}\} < 2 X_M) = p_2.
\end{equation*}
Finally, using the total probability formula, the probability that the first eigenmode is localized on the boundary is given by
\begin{equation*}
P_b = q^2p_1+pqp_2+pqp_3,
\end{equation*}
which gives \eqref{bdprob} in the main text.

\subsection{Calculation of the multimodal probability for the Dirichlet boundary condition}\label{appB2}
Under the Dirichlet boundary condition, the first eigenmode is localized in the longest subregion with $V = 0$ when $K$ is large. If there is a unique longest subregion with $V = 0$, then the first eigenmode is unimodal; if there are two or more longest subregions with $V = 0$, then the first eigenmode is multimodal. Therefore, the event that the first eigenmode is unimodal is equivalent to the event that there is a unique longest subregion with $V = 0$. This probability can be computed explicitly as
\begin{equation*}
\begin{split}
1-P_D &= \sum_{k=1}^{M} \mathbb{P}(\max\{X_1, X_2, \cdots, X_{k-1}, X_{k+1}, \cdots, X_{M}\} < X_k) \\
&= M \; \mathbb{P}(\max\{X_{2}, X_{3}, \cdots, X_{M}\} < X_1) \\
&= M \sum_{n=1}^{\infty} \mathbb{P}(\max\{X_2, X_3, \cdots, X_{M-1}\} < n) \mathbb{P}(X_1 = n) \\
&= M \sum_{n=1}^{\infty} [\mathbb{P}(X_k < n)]^{M-1} \mathbb{P}(X_1 = n)\\
&= M \sum_{n=1}^{\infty} (1 - q^{n-1})^{M-1} q^{n-1} p,
\end{split}
\end{equation*}
which gives \eqref{multiD} in the main text.

\subsection{Calculation of the multimodal probability for the Neumann boundary condition}\label{appB3}
Under the Neumann boundary condition, the first eigenmode is localized in the longest extended subregion with $V = 0$ when $K$ is large. Therefore, the event that the first eigenmode is unimodal is equivalent to the event that there is a unique longest extended subregion with $V = 0$. This event can be classified into the following four situations.

First, we consider the situation where $V(0) = V(1) = 0$, whose probability is $q^2$. In this case, the lengths of extended subregions with $V = 0$ are given by $2 X_1, X_2, \cdots, X_{M-1}, 2 X_M$. Then the probability that there is a unique longest extended subregion is given by
\begin{equation*}
\begin{split}
p_1 = &\; \sum_{k=2}^{M-1} \mathbb{P}(\max\{2 X_1, X_2, \cdots, X_{k-1}, X_{k+1}, \cdots, 2 X_{M}\} < X_k) \\
&\; + \mathbb{P}(\max\{X_2, \cdots, X_{M-1}, 2 X_M\} < 2 X_1) + \mathbb{P}(\max\{2 X_1, X_2, \cdots, X_{M-1}\} < 2 X_M) \\
= &\; (M-2) \mathbb{P}(\max\{2 X_1, X_3 \cdots, 2 X_{M}\} < X_2) + 2 \mathbb{P}(\max\{X_2, \cdots, X_{M-1}, 2 X_M\} < 2 X_1) \\
= &\; (M-2) \sum_{n=1}^{\infty} \mathbb{P}(\max\{2 X_1, \cdots, 2 X_{M}\} < n) \mathbb{P}(X_2 = n) \\
&\; + 2 \sum_{n=1}^{\infty} \mathbb{P}(\max\{X_2, \cdots, 2 X_M\} < 2 n) \mathbb{P}(X_1 = n) \\
= &\; (M-2) \sum_{n=1}^{\infty} [\mathbb{P}(2 X_k < n)]^2 [\mathbb{P}(X_k < n)]^{M-3} \mathbb{P}(X_2 = n) \\
&\; + 2 \sum_{n=1}^{\infty} [\mathbb{P}(X_k < 2 n)]^{M-2} \mathbb{P}(X_k < n) \mathbb{P}(X_1 = n) \\
= &\; (M-2) \sum_{n=1}^{\infty} (1 - q^{[\frac{n-1}{2}]})^2 (1 - q^{n-1})^{M-3} q^{n-1} p + 2 \sum_{n=1}^{\infty} (1 - q^{2n-1})^{M-2} (1 - q^{n-1}) q^{n-1} p.
\end{split}
\end{equation*}
Where $[x]$ represents the largest integer less than or equal to $x$. Second, we consider the situation where $V(0) = 0$ and $V(1) = 1$, whose probability is $p q$. In this case, the length of extended subregions with $V = 0$ are given by $2 X_1, X_2, \cdots, X_M$. Then the probability that there is a unique longest extended subregion is
\begin{equation*}
\begin{split}
p_2 = &\; \sum_{k=2}^{M} \mathbb{P}(\max\{2 X_1, X_2, \cdots, X_{k-1}, X_{k+1}, \cdots, X_{M}\} < X_k) \\
&\; + \mathbb{P}(\max\{X_2, \cdots, X_{M-1}, X_M\} < 2 X_1) \\
= &\; (M-1) \mathbb{P}(\max\{2 X_1, X_3 \cdots, X_{M}\} < X_2) + \mathbb{P}(\max\{X_2, \cdots, X_{M-1}, X_M\} < 2 X_1) \\
= &\; (M-1) \sum_{n=1}^{\infty} \mathbb{P}(\max\{2 X_1, \cdots, X_{M}\} < n) \mathbb{P}(X_2 = n) \\
&\; + \sum_{n=1}^{\infty} \mathbb{P}(\max\{X_2, \cdots, X_M\} < 2 n) \mathbb{P}(X_1 = n) \\
= &\; (M-1) \sum_{n=1}^{\infty} \mathbb{P}(2 X_k < n) [\mathbb{P}(X_k < n)]^{M-2} \mathbb{P}(X_2 = n) + \sum_{n=1}^{\infty} [\mathbb{P}(X_k < 2 n)]^{M-1} \mathbb{P}(X_1 = n) \\
= &\; (M-1) \sum_{n=1}^{\infty} (1 - q^{[\frac{n-1}{2}]}) (1 - q^{n-1})^{M-2} q^{n-1} p + \sum_{n=1}^{\infty} (1 - q^{2n-1})^{M-1} q^{n-1} p.
\end{split}
\end{equation*}
Third, we consider the situation where $V(0) = 1$ and $V(1) = 0$, whose probability is $pq$. In this case, the probability that there is a unique longest extended subregion is given by $p_3 = p_2$. Finally, we consider the situation in which $V(0) = V(1) = 1$, whose probability is $p^2$. In this case, the length of the extended subregions with $V = 0$ are given by $X_1, X_2, \cdots, X_M$. Then the probability that there is a unique longest extended subregion is given by $p_4 = 1-p_D$, where $p_D$ is the multimodal probability for the Dirichlet boundary condition. In summary, the probability that the first eigenmode is unimodal is given by
\begin{equation*}
1-P_N = q^2p_1+pqp_2+pqp_3+p^2p_4,
\end{equation*}
which gives \eqref{multiN} in the main text.

\section{Critical threshold of bifurcation}\label{appC}
In the main text, we have shown that bifurcation occurs when the first eigenvalues of the two subsystems $\tilde{S}_1$ and $\tilde{S}_2$ are equal.

\subsection{Eigenvalues of the subsystem $\tilde{S}_1$}
Here we compute the eigenvalues of the subsystem $\tilde{S}_1$ given in \eqref{tildeS}. For convenience, we define
\begin{equation*}
\alpha = \sqrt{\lambda}, \quad \beta = \sqrt{K - \lambda}, \quad t_0 = L_1 / 2.
\end{equation*}
In the interval $[0, t_0]$, the solution of the subsystem $\tilde{S}_1$ is given by
\begin{equation*}
u(x) = A \sin(\alpha x) + B \cos(\alpha x), \quad u'(x) = A \alpha \cos(\alpha x) - B \alpha \sin(\alpha x),
\end{equation*}
and in the interval $[t_0, 1/2]$, the solution of \eqref{tildeS} is given by
\begin{equation*}
u(x) = C \exp(\beta x) + D \exp(-\beta x), \quad u'(x) = C \beta \exp(\beta x) - D \beta \exp(-\beta x),
\end{equation*}
where $A ,B, C, D$ are four constants to be determined. Under the Neumann boundary condition, due to the continuity of the eigenmodes, we have
\begin{equation*}
\begin{split}
&u'(0) = A \alpha = 0 \\
&u'(1/2) = C \beta \exp(\beta/2) - D \beta \exp(-\beta/2) = 0 \\
&u(t_0) = A \sin(\alpha t_0) + B \cos(\alpha t_0) = C \exp(\beta t_0) + D \exp(-\beta t_0) \\
&u'(t_0) = A \alpha \cos(\alpha t_0) - B \alpha \sin(\alpha t_0) = C \beta \exp(\beta t_0) - D \beta \exp(-\beta t_0)
\end{split},
\end{equation*}
which can be rewritten in matrix form as
\begin{equation*}
\begin{split}
\left[\begin{array}{cccc} \alpha & 0 & 0 & 0\\ 0 & 0 & \beta \mathrm{e}^{\frac{\beta}{2}} & - \beta \mathrm{e}^{-\frac{\beta}{2}}\\ \sin\!\left(\alpha t_0\right) & \cos\!\left(\alpha t_0\right) & - \mathrm{e}^{\beta t_0} & - \mathrm{e}^{- \beta t_0}\\ \alpha \cos\!\left(\alpha t_0\right) & - \alpha \sin\!\left(\alpha t_0\right) & - \beta \mathrm{e}^{\beta t_0} & \beta \mathrm{e}^{- \beta t_0} \end{array}\right]
\left[\begin{array}{c} A \\ B \\ C \\ D \end{array}\right]
=
\left[\begin{array}{c} 0 \\ 0 \\ 0 \\ 0 \end{array}\right].
\end{split}
\end{equation*}
Therefore, $\lambda$ is an eigenvalue of the subsystem $\tilde{S}_1$ if and only if the above coefficient matrix $M$ is nondegenerate, i.e.
\begin{equation*}
\det(M) = \alpha \mathrm{e}^{2 \beta t_0} \sin\!\left(\alpha t_0\right) - \beta \mathrm{e}^{\beta} \cos\!\left(\alpha t_0\right) + \alpha \mathrm{e}^{\beta} \sin\!\left(\alpha t_0\right) + \beta \mathrm{e}^{2 \beta t_0} \cos\!\left(\alpha t_0\right) = 0,
\end{equation*}
which can be rewritten as
\begin{equation*}
D_1(K, \lambda) = \alpha \tan(\alpha t_0) - \beta \tanh(\beta (1/2 - t_0)) = 0.
\end{equation*}

\subsection{Eigenvalues of the subsystem $\tilde{S}_2$}
Here we compute the eigenvalues of the subsystem $\tilde{S}_2$ given in \eqref{tildeS}. For convenience, we define
\begin{equation*}
t_1 = L_4/2, \quad t_2 = L_4/2 + L_3, \quad t_3 = 1/2.
\end{equation*}
In the interval $[0, t_1]$, the solution of the subsystem $\tilde{S}_2$ is given by
\begin{equation*}
u(x) = A \exp(\beta x) + B \exp(-\beta x), \quad u'(x) = A \beta \exp(\beta x) - B \beta \exp(-\beta x),
\end{equation*}
in the interval $[t_1, t_2]$, the solution of the subsystem $\tilde{S}_2$ is given by
\begin{equation*}
u(x) = C \sin(\alpha x) + D \cos(\alpha x), \quad u'(x) = C \alpha \cos(\alpha x) - D \alpha \sin(\alpha x),
\end{equation*}
and in the interval $[t_2, t_3]$,  the solution of the subsystem $\tilde{S}_2$ is given by
\begin{equation*}
u(x) = E \exp(\beta x) + F \exp(-\beta x) \qquad u'(x) = E \beta \exp(\beta x) - F \beta \exp(-\beta x),
\end{equation*}
where $A ,B, C, D, E, F$ are six constants to be determined. Under the Neumann boundary condition, due to the continuity of the eigenmodes, we have
\begin{equation*}
\begin{split}
&u'(0) = A \beta - B \beta = 0 \\
&u(t_1) = A \exp(\beta t_1) + B \exp(-\beta t_1) = C \sin(\alpha t_1) + D \cos(\alpha t_1) \\
&u'(t_1) = A \beta \exp(\beta t_1) - B \beta \exp(-\beta t_1) = C \alpha \cos(\alpha t_1) - D \alpha \sin(\alpha t_1) \\
&u(t_2) = C \sin(\alpha t_2) + D \cos(\alpha t_2) = E \exp(\beta t_2) + F \exp(-\beta t_2) \\
&u'(t_2) = C \alpha \cos(\alpha t_2) - D \alpha \sin(\alpha t_2) = E \beta \exp(\beta t_2) - F \beta \exp(-\beta t_2) \\
&u'(1/2) = E \beta \exp(\beta/2) - F \beta \exp(-\beta/2) = 0,
\end{split}
\end{equation*}
which can be rewritten in matrix form as
\begin{equation*}
\begin{split}
\left[\begin{array}{cccccc} 1 & -1 & 0 & 0 & 0 & 0\\ \mathrm{e}^{\beta t_1} & \mathrm{e}^{- \beta t_1} & - \sin\!\left(\alpha t_1\right) & - \cos\!\left(\alpha t_1\right) & 0 & 0\\ \beta \mathrm{e}^{\beta t_1} & - \beta \mathrm{e}^{- \beta t_1} & - \alpha \cos\!\left(\alpha t_1\right) & \alpha \sin\!\left(\alpha t_1\right) & 0 & 0\\ 0 & 0 & \sin\!\left(\alpha t_2\right) & \cos\!\left(\alpha t_2\right) & - \mathrm{e}^{\beta t_2} & - \mathrm{e}^{- \beta t_2}\\ 0 & 0 & \alpha \cos\!\left(\alpha t_2\right) & - \alpha \sin\!\left(\alpha t_2\right) & - \beta \mathrm{e}^{\beta t_2} & \beta \mathrm{e}^{- \beta t_2}\\ 0 & 0 & 0 & 0 & \mathrm{e}^{\frac{\beta}{2}} & - \mathrm{e}^{-\frac{\beta}{2}} \end{array}\right]
\left[\begin{array}{c} A \\ B \\ C \\ D \\ E \\ F \end{array}\right]
=
\left[\begin{array}{c} 0 \\ 0 \\ 0 \\ 0 \\ 0 \\ 0 \end{array}\right].
\end{split}
\end{equation*}
Therefore, $\lambda$ is an eigenvalue of the subsystem $\tilde{S}_1$ if and only if the above coefficient matrix $M$ is nondegenerate, i.e.
\begin{equation*}
\begin{split}
\det(M) =&\; \alpha^2 \mathrm{e}^{2 \beta (t_1 + t_2)}  + \alpha^2 \mathrm{e}^{2 \beta (t_1 + x_{3})}
+ \beta^2 \mathrm{e}^{2 \beta (t_1 + t_2)} - \beta^2 \mathrm{e}^{2 \beta (t_1 + x_{3})}
+ \alpha^2 \mathrm{e}^{2 \beta t_2}  + \alpha^2 \mathrm{e}^{2 \beta x_{3}} \\
&\; - \beta^2 \mathrm{e}^{2 \beta t_2} + \beta^2 \mathrm{e}^{2 \beta x_{3}}
+ 2 \alpha \beta \mathrm{e}^{2 \beta (t_1 + x_{3})} \cot(\alpha (t_1 - t_2)) - 2 \alpha \beta \mathrm{e}^{2 \beta t_2} \cot(\alpha (t_1 - t_2)) = 0,
\end{split}
\end{equation*}
which can be rewritten as
\begin{equation*}
\begin{split}
D_2(K, \lambda) = &\; (\alpha^2 - \beta^2)(\mathrm{e}^{2 \beta t_2} + \mathrm{e}^{2 \beta (t_1+t_3)}) / (\mathrm{e}^{2 \beta (t_1+t_3)} - \mathrm{e}^{2 \beta t_2}) \\
&\; + (\alpha^2 + \beta^2)(\mathrm{e}^{2 \beta t_3} + \mathrm{e}^{2 \beta (t_1+t_2)}) / (\mathrm{e}^{2 \beta (t_1+t_3)} - \mathrm{e}^{2 \beta t_2}) \\
&\; + 2 \alpha \beta \cot(\alpha (t_1 - t_2)) = 0.
\end{split}
\end{equation*}

\end{appendices}

\setlength{\bibsep}{5pt}
\small\bibliographystyle{nature}

\end{document}